\documentclass[CRPHYS,Unicode,manuscript]{cedram}

\usepackage{color}
\DeclareUnicodeCharacter{2212}{$-$}

\definecolor{seagreen}{rgb}{0.25,0.58,0.27}
\definecolor{aqua}{rgb}{0,0.588,1.0}
\definecolor{strawberry}{rgb}{1.0,0.0,0.5}

\definecolor{blueberry}{rgb}{0.015686275,0.2,1}

\definecolor{orange}{rgb}{1.0,0.5,0}

\graphicspath{{Figures/}} 

\begin{document}

\title{The glass transition in molecules, colloids and grains: universality and specificity}

\author{\firstname{Olivier} \lastname{Dauchot}}
\address{Gulliver UMR CNRS 7083, ESPCI Paris, PSL Research University, 10 rue Vauquelin, 75005 Paris, France}
%
\author{\firstname{Fran\c{c}ois} \lastname{Ladieu}\IsCorresp} 
\address{SPEC, CEA, CNRS, Universit'{e} Paris-Saclay, CEA Saclay Bat 772, F-91191 Gif-sur-Yvette Cedex, France}
%
\author{\firstname{C. Patrick} \lastname{Royall}} 
\addressSameAs{1}{<repeat address 1>}

\begin{abstract} 
We highlight certain key achievements in experimental work on molecular, colloidal and granular glassformers. This short review considers these three classes of experimental systems and focusses largely on the work of the authors and their coworkers and thus is far from exhaustive. Our goal is rather to discuss particular experimental results from these classes and to explore universality and specificity across the broad range of length-- and time--scales they span. We emphasize that a variety of phenomena, not least dynamical heterogeneity, growing lengthscales and a change in structure, albeit subtle, are now well established in these three classes of glassformer. We then review some experimental measurements which depend more specifically on the class of glassformer, such as the Gardner transition and some which have been investigated more in one or two classes than in all, such as configurational entropy and evidence for a dynamical phase transition. We finally put forward some open questions and consider what could be done to fill some of the gaps between theoretical approaches and experiments.
\end{abstract}

\maketitle
\section{Introduction}
\begin{figure}[t!]
\centering
\includegraphics[width = 0.9\columnwidth]{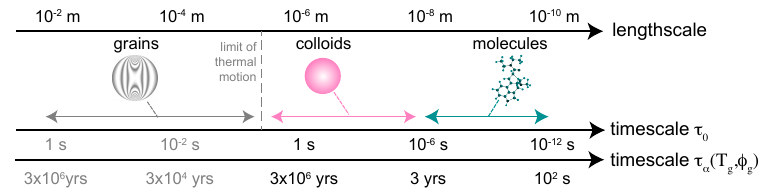}
\caption{\textbf{Classes of glassformers across space and time.} 
Shown as the bottom axes are two corresponding timescales $\tau_0$, the microscopic relaxation timescale pertinent to the high temperature (low packing fraction) case and $\tau_\alpha(T_g,\phi_g)=10^{14}\tau_0$ the alpha relaxation time equivalent to the glass transition of molecules. Note that granular systems are athermal, the timescale depends on the driving and thus is not fixed, here we provide indicative values. The photoelastic disk representing grains is taken from~\cite{thomas2019}.}
\label{figLengthscalesHeart}
\end{figure}

When a molecular liquid is cooled below its melting point sufficiently fast that crystallization is avoided, it enters the meta-stable supercooled regime. Further decreasing the temperature, one observes a dramatic slowing down of the dynamics and a corresponding increase of the structural relaxation time $\tau_\alpha$ over many orders of magnitude. Eventually when $\tau_\alpha$ exceeds 100s, the system is presumed to fall out of equilibrium and to become an amorphous solid, a glass. This is the \emph{molecular} glass transition. The past 30 years have seen a surge of results which have significantly deepened our understanding of the physics of glassforming systems, including several breakthroughs: the emergence of a number of theoretical approaches~\cite{biroli2022,lubchenko2007,parisi2010,chandler2010,speck2019,tarjus2005,janssen2018,charbonneau2005}, advanced computer simulations~\cite{berthier2022}, and the quantitative experimental study of glassy phenomena in very different systems, extending the field of research from molecular liquids to colloidal suspensions and granular assemblies amongst a range of other materials (Fig. \ref{figLengthscalesHeart}) ~\cite{ediger2000,ediger2017,hunter2012,yunker2014,ivlev2012,royall2018jpcm,dauchot2011}.

Exploiting the authors' contributions in molecular (FL), colloidal (CPR) and vibrated granular (OD) glassformers, the present review aims at providing a picture of, what we believe are now, well established experimental results across these material classes. By focussing on our own work, 
we make what may be thought of as a first step towards exploring universal behaviour across these classes of system. We also highlight
phenomena specific to each class of system. We then move on with formulating simple, yet key open questions, in the light of existing theories, when confronted by these results. Finally we discuss some possible strategies to fill the gaps between theoretical approaches and experiments.

We shall largely restrict ourselves to the \emph{approach} to the glass transition in \emph{equilibrated} supercooled liquids 
where, by equilibrated we mean that the observation timescale is much longer than the relaxation time. We refer the interested reader to reviews on ultrastable glasses \cite{ediger2017} and non--equilibrium phenomena such as aging \cite{arceri2020} and shear \cite{bonn2017}. We shall furthermore consider the three classes mentioned above; we refer the interested reader to reviews on oxide \cite{salmon2013}, polymeric \cite{mckenna2017} and metallic \cite{cheng2011} glassformers.

\section{Motivation}
\label{sectionMotivation}
\subsection{Scientific questions opened by the glass transition}
\label{sectionScientificQuestions}

We start by briefly outlining the scientific questions that the experimental studies we review are intended to address, referring the reader to more detailed reviews as appropriate, starting with a general review of the glass transition~\cite{berthier2011}.

It is a basic tenet of materials science that the macroscopic properties of a material are, ultimately, encoded in its structure at the microscopic level. Glassforming systems challenge this viewpoint, as their microscopic structure, when characterized by the pair correlation function, exhibits very little change upon approaching the glass transition; yet they are as solid as any conventional crystalline solid. Although higher--order correlation functions may better characterize the structural evolution in supercooled liquids approaching the glass transition~\cite{royall2015physrep}, there remains the more profound question of what actually leads liquids to solidify without crystallising. Indeed, it has been said that ``there are more theories of the glass transition than there are theorists who propose them''~\cite{chang2008}. 
From the experimental perspective that we pursue here, the challenge boils down to (i) identifying those theoretical approaches which correctly describes the experimental observations and (ii) proposing new experimental observations, which allow for a better discrimination amongst existing theories. 
Here, it is reasonable to separate these theories into two general approaches -- that the glass transition is predominantly driven either by \emph{thermodynamics} or \emph{dynamics}~\cite{cavagna2009}.

The thermodynamic approach to the glass transition pertains to theories, such as the \emph{Random First--Order Transition} (RFOT) theory ~\cite{lubchenko2007}, which invoke a thermodynamic transition towards an ``ideal glass'' state, 
as the underlying cause for the dynamic slowdown. The concept of the ideal glass has its roots in the work of Kauzmann \cite{kauzmann1948}, who predicted that, 
the configurational entropy of supercooled liquids (that part of the entropy which remains after vibrational contributions are excluded) would fall below that of their crystal at a finite temperature $T_k$. In such theories, the relaxation occurs via cooperatively rearranging regions (CRRs). As the system is cooled, their size grows while their timescale increases massively. Approaching $T_k$, the time to relax diverges and the system falls out of equilibrium~\cite{berthier2011,royall2015physrep}.

The alternative scenario, is that the glass transition is a predominantly dynamical phenomenon. One such theory is \emph{Dynamic Facilitation}~\cite{chandler2010,speck2019} which posits that it is a dynamical phase transition between active and inactive \emph{trajectories} that is responsible for the glass transition. In this scenario, the elements of relaxation are short--lived ``pockets'' of mobility, so--called excitations, which are microscopic in time- and length-scale, quite unlike the CRRs mentioned above.

Despite their profound conceptual difference, both scenarios predict a dramatic increase of the relaxation time, as well as dynamic heterogeneity, the phenomenon of different regions of a supercooled liquid exhibiting different relaxation times. Hence the challenge for discriminating amongst them, not withstanding the possibility that the different relaxation mechanisms could hold simultaneously, or even that other mechanism may also exist.

Another phenomenon investigated recently is the \emph{Gardner transition}, originally formulated by Elizabeth Gardner in the context of spin glasses. This is a transition between two kinds of glass, and was predicted from Replica theory that structural glasses could also exhibit this phenomenon \cite{charbonneau2014,charbonneau2017}. We review experimental explorations of the Gardner transition in section \ref{sectionGardner}.

\subsection{A universal slowing down}
\label{sectionClasses}

Besides supercooled liquids, another class of glassforming systems is colloidal suspensions, which exhibit phase behaviour analogous to atomic and molecular systems due to their Brownian motion~\cite{hunter2012,yunker2014}. Here experiments are typically carried out at room temperature, however the interactions between the particles may be changed, for example by the addition of polymer~\cite{likos2001,poon2002}, and colloidal glassforming systems may be formed in an analogous manner to those in molecular systems. Such experiments either move along dilution lines in the phase diagram~\cite{poon2002}, or use constant colloid volume fraction~\cite{royall2018jcp}. The former, while convenient, conserve neither volume nor (osmotic) pressure. The latter may encounter the spinodal line to colloidal liquid--gas phase separation, in which case the system undergoes gelation, a profoundly different scenario of dynamical arrest compared to
vitrification~\cite{royall2018jcp,royall2021jpcm}. An easier way to approach the glass transition in colloidal systems is thus to increase number density, or volume fraction (i.e. packing fraction) $\phi$~\cite{pusey1987}. While some studies have considered charged colloids with a long--range repulsion, which form a so--called Wigner glass~\cite{klix2010,vanderlinden2013}, the vast majority have considered hard sphere--like systems \cite{poon2012,royall2013myth}, and these will form the main focus of our interest. An 
elegant colloidal system to explore glassy behaviour is microgels, whose effective diameter (and thus effective volume fraction) can be controlled \emph{in-situ}, thus enabling an approach to the glass in a single sample, rather than necessitating a new sample for each state point as is the case with conventional hard sphere systems \cite{yunker2014}. Similarly to molecules, crystallisation in colloids must be avoided, often by using a system which is size polydisperse.

The 1990s saw a massive growth in the field of granular matter, leading to work which showed strong experimental evidence of glassy dynamics in dense granular media under low mechanical excitation~\cite{dauchot2011}. The purpose of the mechanical excitation is to supplement for the absence of thermal motion. A major challenge of such experiments is to ensure homogeneous excitation, in order to best mimic Brownian motion and avoid any sort of convective motion. Interactions between particles in such systems are typically close to hard, and therefore they are analogous to colloidal hard sphere--like systems. Here also the system is size polydisperse, often simply bidisperse, and the control parameter of the dynamics is the packing fraction, $\phi$. The majority of the work uses effective two--dimensional systems. These can be compared with work with colloids which can be either 3D or quasi-2D \cite{marcus1996,marcus1999,cui2001,zangi2004,zhang2011prl,gokhale2016}.

\begin{figure}[t!]
\centering
\includegraphics[width = \columnwidth]{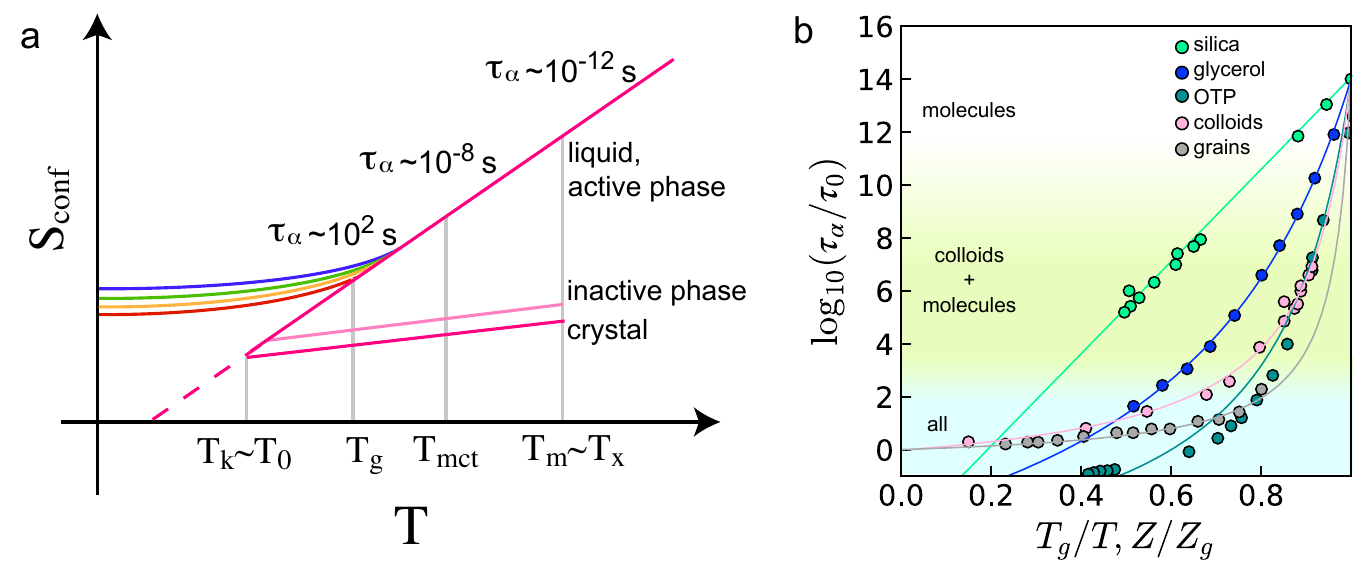}
\caption{\textbf{Roadmaps to the glass transition: the configurational entropy and the ``Angell plot''.} 
(a) The configurational entropy, $S_\mathrm{conf}$, of liquids falls faster than that of crystals as a function of temperature or inverse reduced pressure and  fall below that of the crystal at some temperature $T_k$. $T_g$ is the operational glass transition temperature where the structural relaxation time reaches 100 s in molecular systems. $T_\mathrm{mct}$ is the mode-coupling crossover. $T_m$ is the melting point. The two branches of the dynamical phase transition of dynamic facilitation are indicated. The active phase is effectively indistinguishable from the normal liquid. The inactive phase has a lower configurational entropy, which lies close to, but is presumably slightly larger than, that of the crystal. The lower critical point of the active and inactive phases is thus bounded by the liquid and crystal and lies close to $T_k$. Based on \cite{royall2020jcp}.
(b) In the Angell plot representation the relaxation time is plotted as a function of inverse temperature $1/T$ for molecular systems and reduced pressure $Z$ for colloids and grains. Data for molecular systems silica and and \emph{ortho-}terphenyl (OTP) are taken from~\cite{royall2015physrep}, for glycerol from~\cite{lunkenheimer2002}, for colloids from~\cite{hallett2018} and we computed for the purpose of the present review the data for grains. Data for silica is fitted with an Arrhenius law (a straight line in this representation), other systems are fitted with the Vogel-Fulcher-Tamman expression (Eq. \ref{eqVFT}).
In the case of 3d hard sphere--like colloids, and 
2d hard disc--like grains, the reduced pressure $Z$ is determined from the volume fraction $\phi$ using the Carnahan Starling relation~\cite{hallett2018,royall2017} and the equilibrium equation of state for bidisperse hard discs~\cite{barrio2001} respectively. By convention $T_g$ and $Z_g$ are set such that $\tau_\alpha/\tau_0 = 10^{14}$. Shading denotes the dynamical regimes accessed by the three different classes of glassformer.}
\label{figCavagnaAngellHeartAll}
\end{figure}

As Fig. \ref{figLengthscalesHeart} makes clear, the microscopic timescales pertinent to relaxation in the liquid, be it molecular, colloidal or granular, are massively different. In molecular liquids, the microscopic timescale $\tau_0\simeq 10^{-12}$ s 
corresponds to the thermal motion of the molecules. For colloids, the fundamental timescale is set by the Brownian time to diffuse (say) a diameter
\begin{equation}
\tau_0 = \tau_B = \frac{3\pi\eta\sigma^3}{k_BT}
\label{eqTauB}
\end{equation}
where $\eta$ is the solvent viscosity. Now the size range of colloids in practice is from $10$ nm, to $>1\mu$m, so $\tau_0=\tau_B$ is already $10^3$ to $10^{13}$ larger. Furthermore, the dynamics of the molecules are Newtonian (at a classical level) whilst those of the colloids are Brownian. The case of granular media is even more extreme: millimetre-- or centimetre--sized grains are too big to exhibit significant thermal fluctuations and their erratic motion requires an external source of vibration. As such they obey non equilibrium steady state dynamics. Given these differences, it is all the more fascinating to consider the universal slowing down of the dynamics, emphasized previously for molecular and colloidal systems~\cite{berthier2009pre,gnan2010,royall2015jnonxtalsol} and here with granular systems included, illustrated by the so-called ``Angell plot'' (Fig. \ref{figCavagnaAngellHeartAll})~\cite{angell1988}, where the logarithm of the relaxation time $\tau_\alpha$, scaled by a microscopic time scale $\tau_0$, is plotted as a function of the inverse temperature $T_g/T$, for molecular glassformers, the reduced pressure $Z/Z_g$, for both hard sphere--like colloidal suspensions and mechanically shaken granular assemblies \footnote{Note that in the case of the colloidal systems, the reduced pressure in question is that of the effective one--component colloid system with contributions from smaller components such as microscopic ions integrated out \cite{likos2001}.}. This observation has driven an intense experimental effort in investigating similarities and differences amongst these systems and probing the robustness of various glass forming scenario, by comparing their behavior. Now, if we are to suppose some kind of universality, then the timescale of the operational glass transition $\tau_\alpha(T_g)=10^{14}\tau_0$ should also vary dramatically. 
That is to say, one may introduce a \emph{scaled} relaxation time $\tau_\alpha/\tau_0$ to compare between different classes of glassformer.

In this way, we arrive at $\tau_\alpha^\mathrm{mol}(T_g)=100$ s, $\tau_\alpha^\mathrm{coll}(\phi_g)\sim 10^{14}$ s  and $\tau_\alpha^\mathrm{gran}(\phi_g) \sim 10^{14}$ s for molecules, and micron--sized colloids and grains respectively. As shown in Fig.~\ref{figLengthscalesHeart}, it is important to consider the differences in these timescales in the analysis of experimental data. For example, in Fig.~\ref{figCavagnaAngellHeartAll}(b) we see that while the molecular data indeed covers 14 decades of change in relaxation time, the colloidal and granular data is rather more modest: if the data for micron-sized colloids were also to reach 14 decades of change in relaxation time, then the experiment would need to have run for $\tau_\alpha^\mathrm{col,gran}(T_g)=3$ million years, which is rather challenging to put it mildly! Therefore, any discussion of universality in glassy behaviour needs to consider this massive disparity in absolute timescale. One consequence of this is that in practise, relative relaxation times in experimental studies with micron--sized colloids and grains are at best $\tau_\alpha^\mathrm{coll,gran}(T_g) \sim 10^{7}$.

\section{Dynamical characterisation of the Glass Transition}
\label{sectionCharacterisation}

We now briefly introduce the main dynamical quantities introduced to characterize the dynamics of our three classes of system when approaching the glass transition. We refer the reader to a detailed discussion of these quantities in~\cite{berthier2011}.
From a conceptual point of view, a natural thing to do is to introduce a mobility field 
\begin{equation}
    c({\bf r},t,0) = \sum_i c_i(t,0)\delta({\bf r}-{\bf r}_i),
    \label{eqCmobility}
\end{equation}
where, ${\bf r}_i(t)$ is the position of particle $i$ at time $t$ and $c_i(t,0)$ is a measure of how much it has moved between time $0$ and time $t$. 
The spatial average of the mobility field,
\begin{equation}
    \hat{C}(t,0) = \frac{1}{V}\int d{\bf r} c({\bf r},t,0),
    \label{eqChat}
\end{equation}
describes how much a given configuration of the system has evolved between $0$ and $t$. Further averaging over the realizations, or the thermal statistics, leads to the temporal correlator $C(t)=\langle \hat{C}(t,0)\rangle$ describing the relaxation of the system. This correlator typically exhibits a stretched-exponential decay which, being fitted by the Kohlrausch-Williams-Watts (KWW) law
\begin{equation}
C(t)=c \exp \left(-(t/\tau_{\alpha})^{b} \right)
\label{eqKWW}
\end{equation}
with $0<c\leq1$ and $0<b\leq1$, allows one to obtain the relaxation time $\tau_{\alpha}$.

In order to examine spatial heterogeneities of the dynamics, one considers the spatial correlator of the mobility field:
\begin{equation}
G_4(\bf{r},t) \equiv \langle\delta c(\bf{0},0,t)\delta c(\bf{r},0,t)\rangle   
\label{eqG4}
\end{equation}
which is termed a four-point correlator, because it examines the two point spatial correlation of a field that already is a two times observable. If there is a dominant lengthscale $\xi_4$ controlling the dynamical heterogeneities, which would grow when approaching the glass transition, one expects by analogy with critical phenomena that 
\begin{equation}
G_4(r,t) \simeq \frac{1}{r^p}e^{-r/\xi_4(t)}. 
\label{eq:G4}
\end{equation}
where $p$ is some critical exponent, i.e. where $G_4$ is scale invariant up to $r \approx \xi_4$. It is also natural to define the susceptibility associated with such spatial correlations:
\begin{equation}
\chi_4(t) \equiv \rho \int dr G_4(r,t)    
\end{equation}
which is easily related to the fluctuations of $\hat{C}$:
\begin{equation}
\chi_4(t) = N \langle \delta\hat{C}^2 \rangle.    
\end{equation}
The above definitions follow exactly those used to describe standard critical phenomena, except that the field of interest here is a mobility, instead of a static quantity such as, for instance, the local magnetization. There is thus no surprise, at least for molecular liquids and colloidal suspensions, in the fact that the susceptibility can also be computed as a response function, a point of significant importance for 
experimental purposes. As a matter of fact, it was demonstrated experimentally that it is also the case for the mechanically agitated grains~\cite{lechenault2008a}.

The choice of the mobility field naturally depends on the experimental conditions. 
Typically a scattering experiment would consider $c_i(t,0)=\exp \left( i {\bf q}({\bf r}_i(t)-{\bf r}_i(0)) \right)$, 
while in real space one usually prefers to use $c_i(t,0)=\exp \left(-({\bf r}_i(t)-{\bf r}_i(0))/a \right)^2$
where $a$ or $2\pi /q$ is the microscopic length characterizing a relaxation event at the particle scale and is of the order of a fraction of the particle diameter or interaction range. It is then straightforward to relate $C(t)$ with the self intermediate scattering function, that is the Fourier transform of the distribution of the particle displacements, also called the self van Hove function:
\begin{equation}
F_s(\mathbf{q},t)=\frac{1}{\rho}\langle \rho_{{\bf q}}(t)\rho_{-{\bf q}}(0)\rangle=\int d{\bf r}\; G_s({\bf r},t)\exp\left(-i{\bf q}\cdot{\bf r} \right)
\label{eqISF}
\end{equation}
Similarly, for the second choice of $c_i$, $C(t)$ is related to the self part of the so called overlap function:
\begin{equation}
Q(a,t)=\int d\mathbf{r}d\mathbf{r'}\langle\rho(\mathbf{r},t)w_a(\mathbf{r}-\mathbf{r'})\rho(\mathbf{r'},0)\rangle,
\end{equation}
with $w_a({\bf r-r'})=e^{-({\bf r-r'})^2/a^2}$.

In all cases, for a given temperature $T$, or packing fraction $\phi$, the correlations, hence the susceptibility, attain a maximum at a certain $t=\tau_{h}\simeq \tau_{\alpha}$, and then die away. The maximum of $\chi_4$ corresponds to correlated relaxations involving a number $N_\mathrm{corr,4}$ of particles, where $N_\mathrm{corr,4} \propto \mathrm{Max}_{t} (\chi_4(t))$. One should however keep in mind that $\hat{C}$, and hence $\chi_4$, depend on the probe length $a$ or $2\pi /q$ and in some cases a careful analysis of this dependence is necessary~\cite{dauchot2005,lechenault2008b}. 
Furthermore the knowledge of the prefactor $A(t)$ in eq.~(\ref{eq:G4}) is necessary, if one is to obtain absolute values for $N_\mathrm{corr,4}$.

\section{Measurement techniques}

In this short review, we will consider three experimental techniques that can be used to access the quantities described in the previous section and that are pertinent to the results we shall discuss: real space imaging, light scattering, and dielectric response. Real space imaging typically applies to micron-sized colloids and 2d granular assemblies, light scattering applies to sub-micron colloidal suspensions and molecular liquids, adapting the light wavelength to the size of the constituents, and dielectric response applies to molecular liquid. We refer to Ref. ~\cite{alba2022} for a description of light scattering applied to molecular liquids and concentrate here on the other techniques.

\subsection{Real space imaging in colloidal suspensions and granular assemblies}
\label{sectionReal} 
Whether it be through the use of microscopy or macroscopic imaging techniques, real space imaging gives access to the positions ${\bf r}_i$ of the individual colloids or grains at a certain frame rate, which is specific to the system and state point of interest, so--called particle--resolved studies (PRS) \cite{ivlev2012}. Then tracking algorithms allow one to reconstruct the particle trajectories ${\bf r}_i(t)$ and one ends up with experimental data~\cite{ivlev2012,leocmach2013sm} akin to those obtained from molecular simulations, albeit with the addition of experimental uncertainties.

These uncertainties take a variety of forms. First is the tracking error in the coordinate position. While (due to their larger size), this is just 1\% of the diameter or even less in the case of grains, for colloids this is typically 5\% of the diameter. One may need to add errors to coordinates obtained from simulation to obtain agreement with experimental data \cite{ivlev2012,royall2007jcp,pinchaipat2017}. Other issues include missing particles and ``ghost particles'' where the tracking algorithm erroneously identifies a particle which is not present in the experiment. These missing particles and ``ghost particles'' can account for up to 5\% of the total particles in poorer quality data.

\begin{figure}[t]
\centering
\includegraphics[width =1.0\columnwidth]{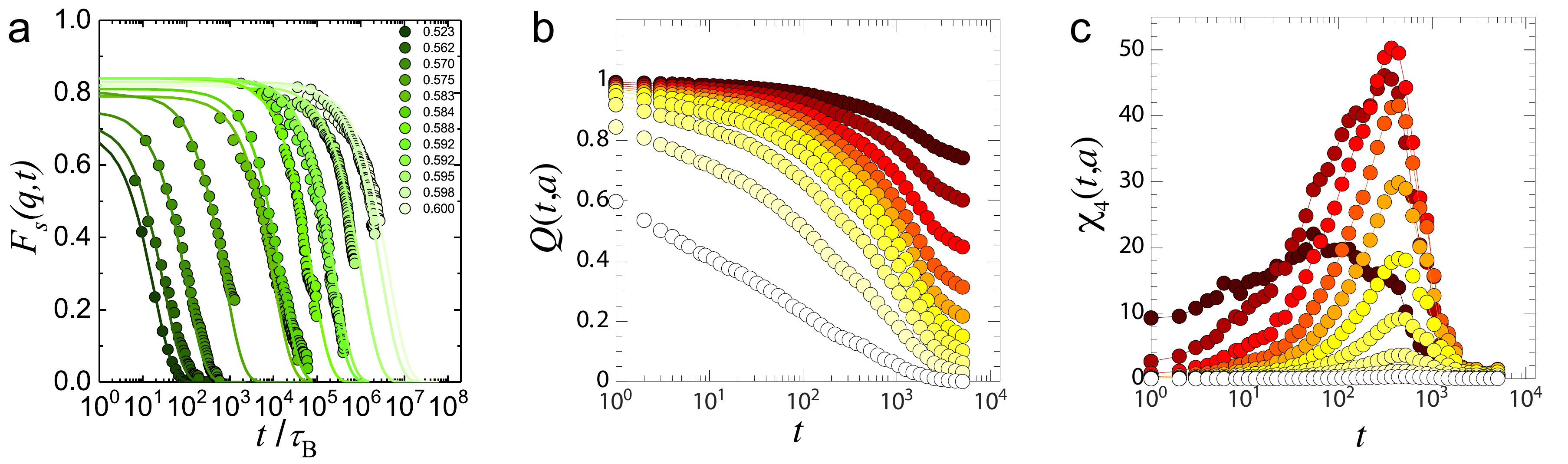}
\caption{\textbf{Measuring relaxation and dynamic heterogeneity in real space from co--ordinate data.}
(a) Intermediate scattering functions for a colloidal system at the volume fractions listed. The wavevector is close to the first peak of the static structure factor. Reproduced from \cite{hallett2018}.
(b) Relaxation curves and (c) Dynamical heterogeneities evaluated via $\chi_4$, extracted from real space imaging of a granular experiment at one given packing fraction, colors correspond to different choice of the probing length $a$; there is one length-scale for which dynamical heterogeneities are best captured. Adapted from \cite{dauchot2005}.
}
\label{figISF}
\end{figure}

From the instantaneous position of the particles, one readily computes structural quantities, such as the pair correlation function, or any local structural order parameter, characterizing for instance the hexatic order in 2d or higher--order structures in 3d via bond--orientational order parameters~\cite{steinhardt1983} or geometric motifs thought to be pertinent to the glass transition ~\cite{malins2013tcc}, and their spatial correlations. One can also extract the dynamical properties of the system as a direct implementation of the methods described in section~\ref{sectionCharacterisation} to obtain $Q(a,t)$ and $\chi_4(a,t)$ as illustrated in Fig.~\ref{figISF} We refer the reader to~\cite{cipelleti2011,dauchot2011b} for more details.

Particle--resolved studies of colloidal requires particles of around 2-3 $\mu$m in diameter, while light scattering can use  particles of 200 nm diameter or less. Inspection of Eq. \ref{eqTauB} and Fig. \ref{figLengthscalesHeart} indicated that using smaller colloids can lead to much higher $\tau_\alpha/\tau_0$ for a given measurement time. \emph{Nano--}particle resolved studies uses recent developments in super--resolution microscopy \cite{hell2014} to image particles around ten times smaller in diameter than conventional PRS which allows access to values of  $\tau_\alpha/\tau_0$ around 1000 times larger than conventional approaches.

\subsection{Light scattering in colloidal suspensions}
\label{sectionLight}
The majority of the light scattering work on colloidal glassformers has used dynamic light scattering (DLS), and this will form the focus of our discussion. We refer the reader to~\cite{chu1991} for a more detailed discussion and to~\cite{cipelleti2011} for a precise exposure to the subtleties of DLS, which we here briefly summarize for the purpose of completeness.

DLS probes the temporal fluctuations of the refractive index of the colloidal suspension, where the scattering occurs because of the mismatch between the index of refraction of particles and that of the solvent. A central advantage of DLS is that it probes a very large number of particles simultaneously, yielding very good statistics. 
Experimentally, one measures the temporal correlations of the light intensity $I(t)$ scattered at a wave vector $q = 4\pi /\lambda \sin(\theta/2)$, where $\theta$ is the scattering angle and $\lambda$ is the wavelength in the solvent of the incoming laser beam:
\begin{equation}
g_2(t) = \frac{\langle I(t)I(0) \rangle}{\langle I(0) \rangle^2},
\end{equation}
where the average is a temporal average, assuming ergodicity, a condition which becomes eventually difficult to ensure close to the glass transition.
Assuming single scattering, the intermediate scattering function (ISF) simply relates to $g_2$:
\begin{equation}
\label{eq:F}
F(q,\tau) \equiv \frac{1}{N} \left\langle \sum_{j,k} \exp\left\{-i
\mathbf{q}\cdot[\mathbf{r}_j(t+\tau)-\mathbf{r}_k(t)]\right\} \right\rangle = \sqrt{c [g_2(\tau)-1]},
\end{equation}
where $c \ge 1$ depends on the optics. One notes that DLS provides access to the full ISF, but it is possible to reduce it to its self part by making the contribution of the $j \ne k$ terms vanish from Eq.~(\ref{eq:F}). To enrich the temporal average, which can be limited, especially close to the colloidal glass transition, it is common to use a camera as a multi-pixel detector and also average over the different pixels, which correspond to statistically independent speckle.

In the above procedure, the light intensity is averaged over both space and time and no information can be extracted on dynamical heterogeneity. One way to go further is the so-called Time Resolved Correlation (TRC)~\cite{cipelletti2002}, where the scattered light intensity is averaged over the pixels only, leading to 
\begin{equation}
\label{eq:cI} c_I(t,0)= \frac{\langle I_p(t)I_p(0) \rangle_p}{\langle I_p(t) \rangle_p \langle I_p(0)\rangle_p}-1,
\end{equation}
a formal equivalent of $\hat{C}(t,0)$, and the usual intensity correlation function $g_2(\tau)-1$
is the temporal average of $c_I(t,\tau)$, exactly as $C(t) = \langle \hat{C}(t,0) \rangle$ (Eqs. \ref{eqCmobility} and \ref{eqChat}).
Having access to $\hat{C}(t,0)$, one extracts the dynamical susceptibility by computing its temporal variance, following section~\ref{sectionCharacterisation}.
The second way to access spatial correlations is Photon Correlation Imaging (PCI)~\cite{duri2009}. Here spatial resolution results from the modification of the collecting optics, using a diaphragm to limit the range of $q$ vectors accepted by the detector, in such a way that each pixel of the sensor is illuminated by light issued from a small region of
the sample and scattered in a small solid angle associated with the same, well defined scattering vector. Dividing the images in small sub-regions, one then accesses directly to a local measure of the temporal decorrelation, which serves as the mobility field to compute $G_4$ or $\chi_4$.

\subsection{\label{SectionDielTech} Dielectric response in molecular liquids}
In molecular liquids, except in 
special cases \cite{huang2013}, it is impossible to perform real space imaging to reconstruct particle trajectories as in section \ref{sectionReal}. Moreover single-component liquids can be obtained with a high degree of purity, yielding the chemical uniformity to be so good that the very existence of dynamic heterogeneity (DH) is far from obvious \emph{a priori}. However, in the 1990s, 
several kinds of sophisticated experimental techniques established that there exists a significant fraction of molecules relaxing more slowly than the average, i.e. that there is 
a distribution of molecular relaxation times in single component liquids. We refer the reader to  Ref. \cite{richert2011} for a review, and for the detailed explanations about the fact that most of these experiments prove the heterogeneous character of the dynamics in time but cannot access the associated spatial scale. The two exceptions to this lack of spatial information about DH's are two experimental \emph{tour de force}, namely  4D-NMR \cite{tracht1998} and Atomic Force Microscopy experiments \cite{vidal2000}, where the typical size of DH's was estimated to be a few molecular diameters \cite{tracht1998} close to $T_g$ with a trend to increase upon cooling.

However, less specific methods are needed for a systematic study of DH. Two main methods were elaborated based either on \textit{linear} responses or on \textit{nonlinear} responses. We shall briefly explain each 
by using the language of dielectric response $\epsilon \equiv 1+\chi$, where one monitors the polarisation $P= \epsilon_0 \chi E$ stemming from the rotation of molecules submitted to an external electric field $E$ of angular frequency $\omega$. Here $\epsilon_0$ is the vacuum dielectric constant.

\begin{figure}[t!]
\centering
\includegraphics[width = \columnwidth]{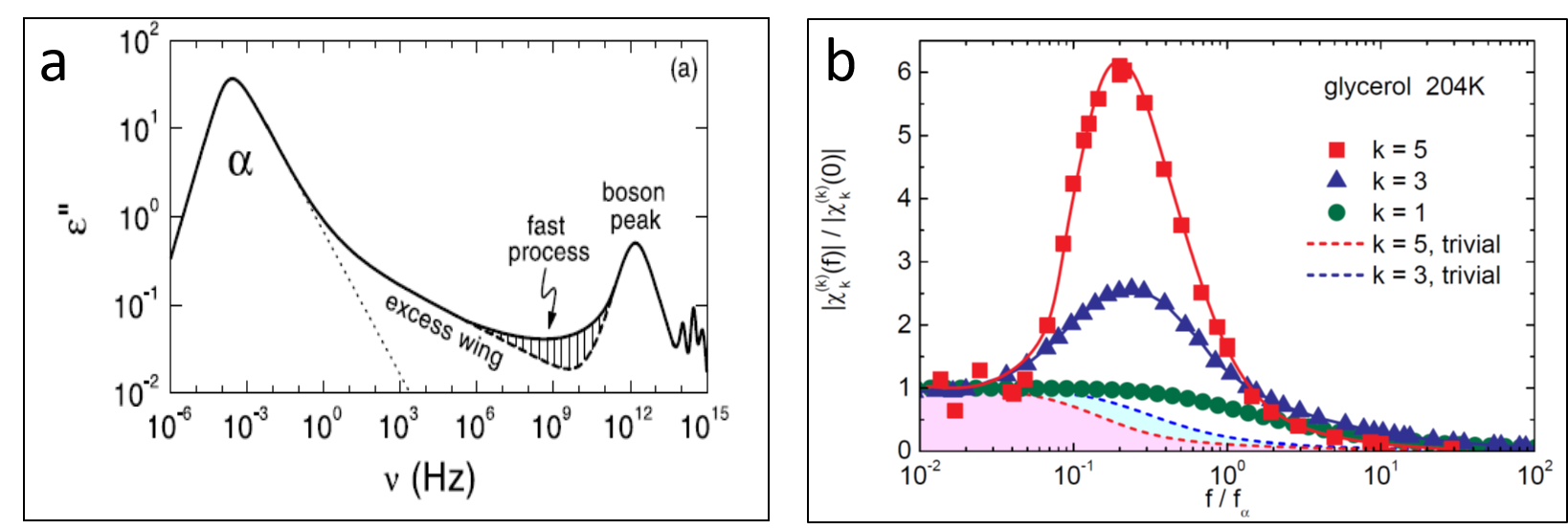}
\caption{ {\bf Linear and nonlinear dielectric spectra in molecular supercooled liquids}. 
(a) Adapted from \cite{lunkenheimer2002}: Typical spectrum for the imaginary part of the linear dielectric susceptibility. 
(b) Adapted from \cite{albert2016}: Modulii of the linear ($k=1$), third order ($k=3$) and fifth order ($k=5$) susceptibility in glycerol at $204$K -which amounts to $T/T_g \simeq 1.08$ since in glycerol $T_g \simeq 188$K-. The trivial case of an ideal gas of non interacting molecules yields monotonically  
decreasing responses when increasing frequency, at all order $k$ in the field. }
\label{figMeasdiel}
\end{figure}

Figure \ref{figMeasdiel}(a) shows a typical spectrum that one obtains for the imaginary part $\epsilon'' \equiv \chi_1''$ where $\chi_1$ is the \textit{linear} complex dielectric susceptibility. We shall not discuss in detail the features appearing at high (excess wing or $\beta$ peak) or very high frequencies (Boson peak) and shall focus on the $\alpha$ relaxation showing up as a peak around the frequency $f_\alpha$ related to the typical molecular relaxation time $\tau_\alpha =  1/(2\pi f_{\alpha})$. With respect to the ideal case of non interacting molecules, the $\alpha$ peak in supercooled liquids is asymmetric in frequency, which has been interpreted as an indication that a large distribution of relaxation times $\tau$ exists at each given temperature $T$. Let us stress that this distribution is not a proof of DH; it could be compatible with 
relaxation being the same everywhere in the liquid and occurring as a stretched exponential in time \cite{richert2011}. However the temperature dependence $\tau_{\alpha} (T)$ can be related to DH's since it was theoretically established \cite{dalle2007,berthier2007a,berthier2007b} that $T \chi_T \propto \vert \partial \ln{ \tau_{\alpha}}/(\partial \ln T) \vert $ is an estimator of $N_\mathrm{corr}(T)$. Physically this comes from the fact that the larger $N_\mathrm{corr}$, the higher the activation energy, and the longer $\tau_{\alpha}$.

The second method involves \textit{non linear} dielectric susceptibilities, for example the third order 
$\chi_3$ defined as 
\begin{equation}
\chi_3  \equiv \lim_{\mathrm{E \to 0}} \left[\frac{P/\epsilon_0 - \chi_1 E}{E^3}\right].     
\label{eqChi3}
\end{equation}
Contrary to its linear counterpart $\chi_{1}(\omega)$, the nonlinear spectrum $\chi_3(\omega)$ is \textit{qualitatively} different in supercooled liquids and in an ideal non interacting gas of dipoles as is taken to be the case for a normal liquid. This is illustrated in Fig. \ref{figMeasdiel}(b) where the moduli of the responses of order $k$ in the field are compared for supercooled glycerol: the linear $k=1$ spectrum varies monotonically in frequency, contrarily to nonlinear responses which exhibit a hump in frequency, which increases with $k$. Physically, this can be understood by considering an amorphously ordered domain containing $N_\mathrm{corr}$ molecules. Since the supercooled liquid is assumed to be comprised of  independent domains, one obtains its polarisation by adapting the formula valid for an ideal gas $P/\epsilon_0 = (\mu/a^3) L(\mu E/(k_BT))$ where $L$ is the Langevin function, $a^3$ the molecular volume, and $\mu$ the molecular dipole. Using this formula for each domain amounts to the two following renormalisations: $a^3 \to N_\mathrm{corr} a^3$ and $\mu \to \sqrt{N_\mathrm{corr}}\mu$, where the square root comes from the amorphous nature of the order, yielding molecular dipoles pointing in (seemingly) random directions. Plugging this into the expansion of the Langevin function directly yields $\chi_k \propto N_\mathrm{corr}^{(k-1)/2}$, which explains the qualitative difference between the linear spectrum (which is blind to amorphous ordering) and the non linear spectra, see for example \cite{albert2016,gadige2017,biroli2021} for detailed explanations. In particular, these references compare the temperature dependence of $\chi_5$ and of $\chi_3$ from which one deduces that the amorphously ordered domains close to $T_g$  are compact objects, i.e. their fractal dimension $d_f$ is found to be that of the embedding space. Though experimentally challenging, nonlinear responses have been achieved by several teams \cite{crauste2010,bauer2013,casalini2015,young2017} on various liquids, and even at high pressures \cite{casalini2015}.

\section{A universal slowing down}
\label{sectionUnivslow}

We begin our analysis of the results of experiments on molecular, colloidal and granular systems by considering the most basic property of glassforming systems, the relaxation time. Figure \ref{figCavagnaAngellHeartAll}(b) shows that the relaxation time of all three classes of interest may be fit with the VFT expression
\begin{equation}
\tau_{\alpha}=\tau_{0}\exp\left[ \frac{D}{T-T_{0}} \right],
\label{eqVFT}
\end{equation}
with $T$ being replaced by $1/Z$ in the cases of colloids and granular media.
$T_{0}~(\simeq T_{{K}}$) is rather lower than the experimental glass transition temperature $T_{{g}}$ and $D$ is a measure of the \textit{fragility} -- the degree to which the relaxation time increases as $T_g$ is approached. Note that the case of Arrhenius dependence (e.g. of silica and other ``strong'' glasses) is rather poorly described by (\ref{eqVFT}).

\begin{figure}[b!]
\centering
\includegraphics[width =0.95\columnwidth]{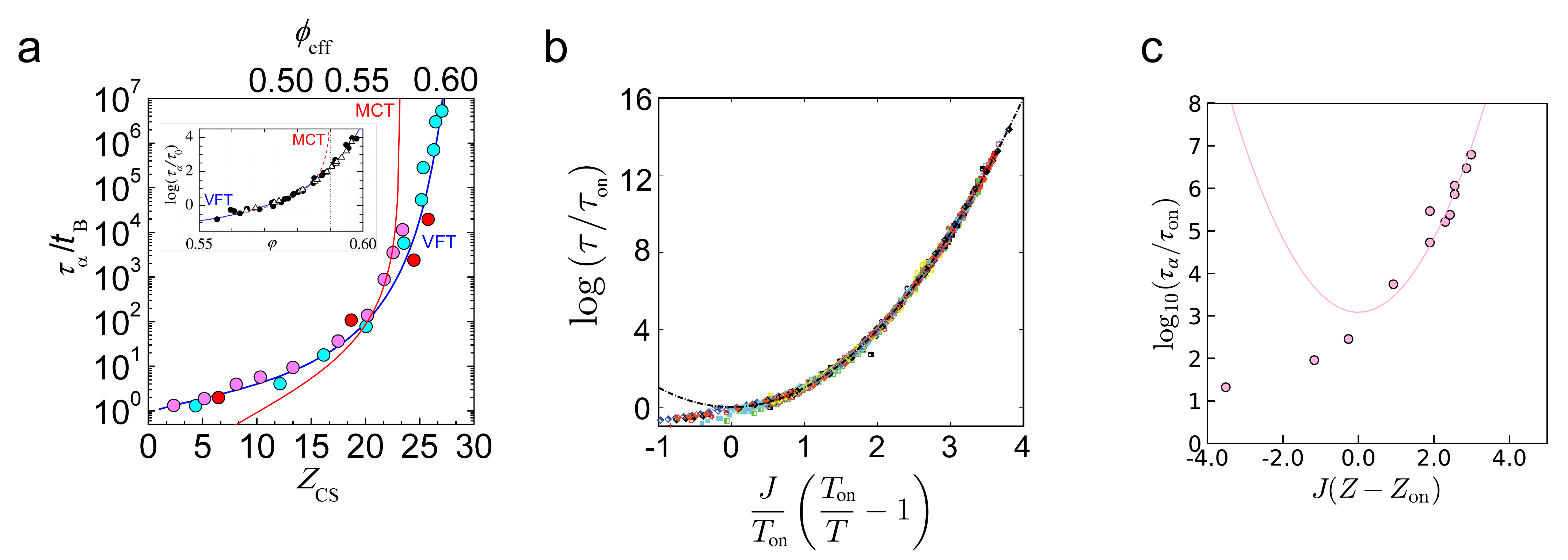}
\caption{\textbf{Experimental tests of alternative theoretical descriptions of the increase in relaxation time.}
(a) Relaxation beyond the mode--coupling crossover in colloidal experiments. Main panel: real space measurements with hard sphere like colloids (pink data are computer simulation). Reproduced from \cite{hallett2018}.
Inset: Light scattering measurements, reproduced from \cite{brambilla2009}. Both these data sets show relaxation past the MCT crossover ($\phi_\mathrm{mct}\approx0.58$) as shown by the blue curves. Fits with the VFT expression (Eq. \ref{eqVFT}) provide much better agreement. 
(b) The parabolic law of Dynamic Facilitation theory Eq. \ref{eqParabolic} fitted to a wide range of molecular glassformers. Reproduced from \cite{elmatad2009}.
(c) The parabolic law fitted to data for hard sphere like colloids tracked in real space. Data from \cite{hallett2018}.
}
\label{figParabolicMCT}
\end{figure}

In the case of colloidal systems, earlier work found a good agreement with mode--coupling theory \cite{vanmegen1998,pham2002} leading to a general acceptance of MCT in the colloid community. However, as shown in Fig. \ref{figParabolicMCT}(a), later work which accessed a greater change in scaled relaxation time $\tau_\alpha / \tau_0$ found that, like molecular glassformers, colloidal systems do in fact relax if supercooled past the ``transition'' of standard MCT~\cite{brambilla2009,hallett2018,hallett2020}. On the theoretical side, further developments of MCT could in principle address this issue~\cite{janssen2018,luo2022}. The VFT fit can be rationalized from the thermodynamic approaches to the glass transition (see section \ref{sectionScientificQuestions} and the review of Cavagna  ~\cite{cavagna2009} for an interpretation of $T_0 \approx T_K$) and therefore support the RFOT scenario.  Dynamic faciliation (DF) theory offers an alternative description with a prediction for the relaxation obeying the so--called parabolic law:
\begin{equation}
\tau_{\alpha} = \tau_\mathrm{on}  \exp \left\{  J^2\left(\frac{1}{T}-\frac{1}{T_{\rm on}}\right)^2 \right\}
\label{eqParabolic}
\end{equation}
where $\tau_\mathrm{on}$ is the relaxation time at the onset temperature $T_\mathrm{on}$ of glassy dynamics. $J$ is a system--specific ``coupling'' parameter related to the spin models that DF is inspired by \cite{chandler2010}. One sees in Fig. \ref{figParabolicMCT}(b,c) that the parabolic law provides a fit to the experimental data as good as that of the VFT equation both in molecular and colloidal systems (with $T$ replaced by $1/\phi$ in the case of the latter). The mechanically shaken granular assemblies do not explore enough decades of change in the relaxation time to be of relevance here. In other words, existing experimental data about the relaxation time cannot discriminate amongst conceptually incompatible theoretical approaches. For a more detailed analysis of molecular glassformers, see \cite{hecksler2008,keys2013}.

\section{Dynamical heterogeneities}
\label{sectionDynamical}
\begin{figure}[t!]
\centering
\includegraphics[width = 0.8\columnwidth]{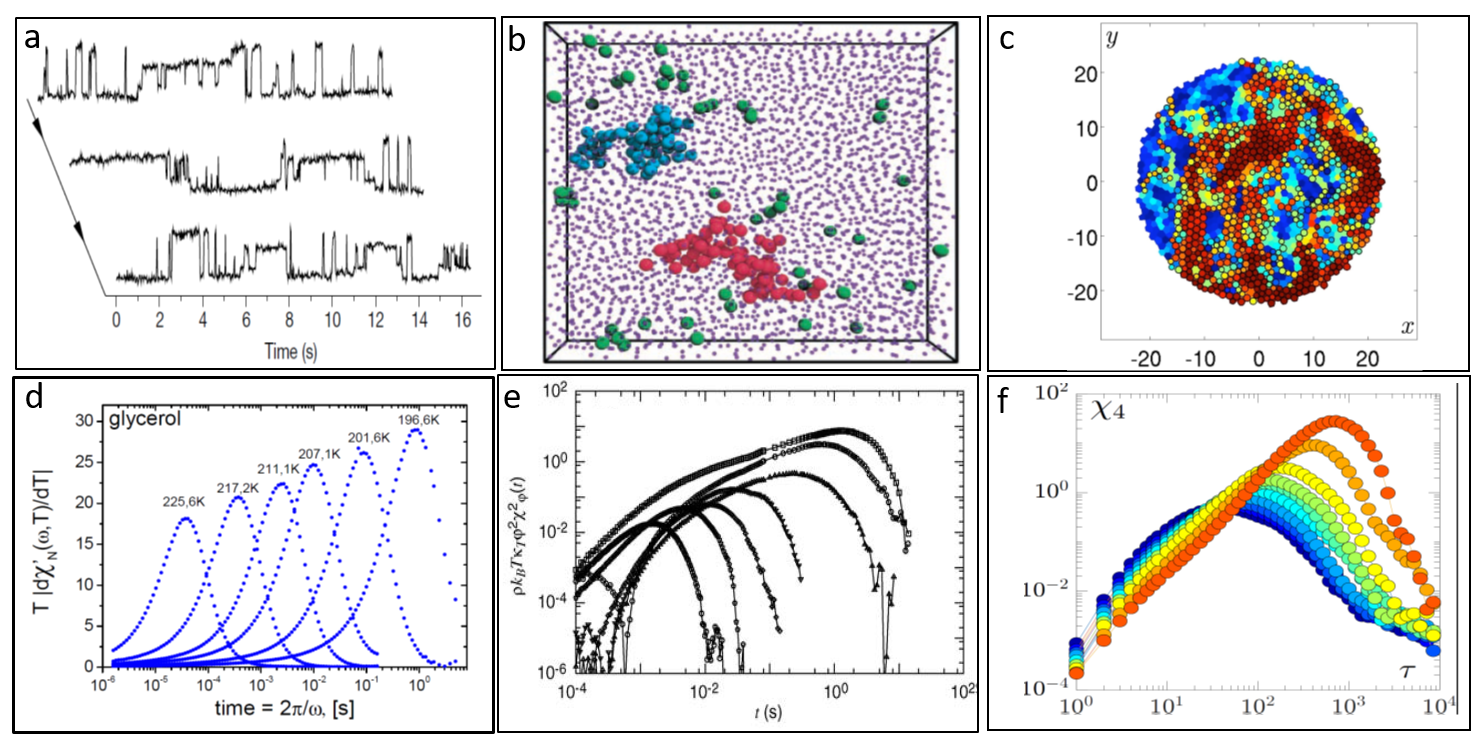}
\caption{{\bf Local and global characterization of Dynamical Heterogeneities (DH's) in molecular, colloidal and granular glass forming systems}. 
Upper (resp. lower) row: local (resp. global) measurements of DH's in (a,d) a molecular glass~\cite{vidal2000,dalle2007}, (b,e) a colloidal glassformer~\cite{weeks2000,berthier2005}, 
(c,f) a vibrated granular glassformer~\cite{keys2007,candelier2010b}. 
Adapted from \cite{vidal2000,dalle2007,weeks2000,berthier2005,candelier2010b}.
}  
\label{figDynhet}
\end{figure}

Dynamical heterogeniety, in which some regions of supercooled liquids exhibit much fast relaxation than others was discovered in the 1990s \cite{perera1996}. This motivated a surge of interest for a quantitative analysis of dynamical heterogeneities. The first work to directly identify DH in colloids that we are aware of was that of Rice and coworkers, who used a 2d system to explore string--like motion similar to early computer simulations and also explored the shape of dynamically heterogeneous regions \cite{marcus1996,marcus1996,cui2001,zangi2004}. Figure \ref{figDynhet} shows (upper row) some DH's observed in (a) a molecular glass (AFM dielectric spectroscopy on a PVac film, \cite{vidal2000}), (b) a colloidal system (confocal microscopy, \cite{weeks2000}), and (c) a granular system (\cite{candelier2009}). In each of these three systems, one directly observes that the dynamics is very heterogeneous both in space and in time. 

As explained above, in granular experiments, the dynamic susceptibility $\chi_4(t)$ can be directly monitored, and one finds, see Fig. \ref{figDynhet}(f), that its maximum in time increases with the area fraction $\phi$. In other kinds of glassforming systems, because measuring $\chi_4(t)$ is either difficult (colloids) or impossible (molecular glassformers), earlier works \cite{berthier2005,berthier2007a, berthier2007b, dalle2007} have used proxies $W(t)$ of $\chi_4(t)$, as illustrated in Fig. \ref{figDynhet}(d,e). More precisely in colloids, one has used in Ref. \cite{berthier2005}  $W_\mathrm{col}(t) \equiv <\rho> k_B T (\phi \partial F_s(q_0,t)/\partial \phi)^2$ where $F_s(q_0,t)$ is the self part of the scattering function taken at the first peak $q_0$; while in molecular liquids a good proxy is 
\begin{equation}
W_\mathrm{mol}(t) \equiv \sqrt{\frac{k_B}{\Delta C_p}} \left| T \frac{\partial \chi_N}{\partial T} \right|  
\label{eqWmol} 
\end{equation}
where $\Delta C_p$ is the jump of the molecular specific heat at $T_g$, 
\begin{equation}
\chi_N = \frac{\chi_1(\omega)}{\chi_1(\omega=0)}
\label{eqChiN}
\end{equation}
and time $t=1/\omega$. Figure \ref{figDynhet}(d,e,f) illustrates that in the three kinds of glassforming systems, the maximum with respect to  
time of $\chi_4(t)$, or of its proxies $W_\mathrm{col}$, $W_\mathrm{mol}$, does increase as one approaches the glass transition: this means that the size of the DH increases as their relaxation time (tremendously) increases to the macroscopic value of $100$s at the glass transition. 

\begin{figure}[t!]
\vspace{-0.0cm}
\centering
\includegraphics[width = 0.8\columnwidth]{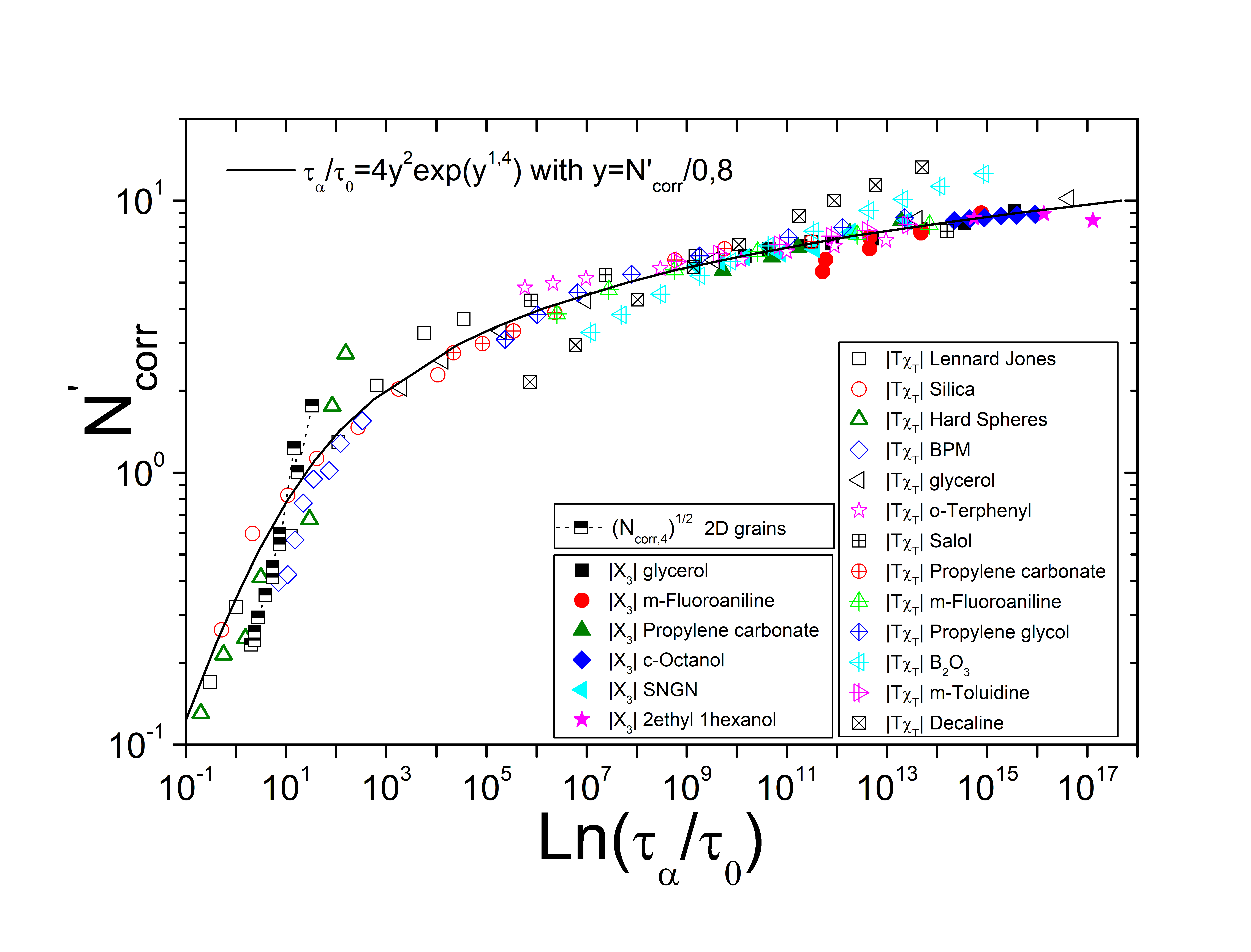}
\vspace{-0.0cm}
\caption{{\bf Comparison of the behavior of $N'_\mathrm{corr}$ evaluated in various glassformers and/or various techniques}. Adapted from Ref. \cite{dalle2007}, \cite{michl2016} and \cite{keys2007,candelier2010b}.  Increase of $N'_\mathrm{corr}$ when approaching the glass transition (see text for details on the normalization and the precise quantities described by $N'_\mathrm{corr}$). This plot combines $2d$ grains and a colloidal system (``Hard Spheres''), as well as $12$ molecular liquids where one has set either (open symbols) $N'_\mathrm{corr} \propto \vert T\chi_T \vert$ or -full symbols- $N'_\mathrm{corr} \propto \vert X_3^\mathrm{max} \vert$ with $\vert X_3^\mathrm{max} \vert$ the maximum of the hump of the (dimensionless) third order susceptibility (see refs \cite{albert2016,gadige2017,biroli2021} for the precise definitions). Two numerical studies (Lennard-Jones and BKS Silica) are also included.
}
\label{figNcorr}
\end{figure}

Beyond this important qualitative result, a quantitative comparison between molecular, colloidal and granular glassforming systems requires to take care of the fact that the conserved quantities are not the same in these three systems. This subtle point has been carefully investigated theoretically \cite{berthier2007a,berthier2007b} with the important outcome that the derivative with respect to $\phi$ has to be squared in $W_\mathrm{col}(t)$ contrary to its counterpart $\partial/\partial T$ in $W_\mathrm{mol}(t)$. This is why in Fig. \ref{figNcorr} where we compare molecular, colloidal and granular systems we shall essentially compare $\mathrm{Max}_t W_\mathrm{mol}$ for molecular glassformers, $\sqrt{\mathrm{Max}_t W_\mathrm{col}}$ for colloids, and $\sqrt{\mathrm{Max}_t \chi_4}$ for granular systems. Defining 
\begin{equation}
\vert T \chi_T \vert \equiv \sqrt{\frac{k_B}{ \Delta C_p}} 
\left| \frac{ T \partial \ln{\tau_{\alpha}}}{ \partial \ln{T}} \right|
\label{eqTchiT}
\end{equation}
one readily shows \cite{dalle2007} that $\mathrm{Max}_t W_\mathrm{mol}$ is proportional to $\vert T \chi_T \vert$, with a prefactor of order $1$. The latter can be dropped since one cannot be sure of the normalisation prefactor when deriving $\chi_4$ and its proxies. This is why, in Fig. \ref{figNcorr}, so as to achieve the best possible collapse \cite{dalle2007} we have chosen, for each system, two multiplicative prefactors, one for $\tau_0$ and another one for $N_\mathrm{corr}$, which is thus relabelled $N'_\mathrm{corr}$. For most systems this factor is close to unity (in any case it is smaller than $10$). Considering the extreme variety of the systems plotted in Fig. \ref{figNcorr}, the quality of the collapse is impressive and suggests that some of the universal aspects of the glass transition is captured with this set of data. Finally, the fact that the points drawn from nonlinear susceptibility (full symbols) superimpose on the trend emerging from the $\vert T \chi_T \vert $ analysis is a very strong check of internal consistency of all these data-sets.

We emphasize that for molecular systems the absolute value of $N_\mathrm{corr}$ is not known because of the normalisation prefactor issue discussed above \cite{dalle2007}. This contrasts with granular~\cite{lechenault2008a} and colloidal systems~\cite{hallett2018,hallett2020}, where a direct measure of the dynamical correlation length can now  be obtained, as shown in Fig.~\ref{figJames}(c) in the case of colloidal glassformers.

\section{Structure}
\label{sectionStructure}

\begin{figure}[b!]
\centering
\includegraphics[width = 0.95\columnwidth]{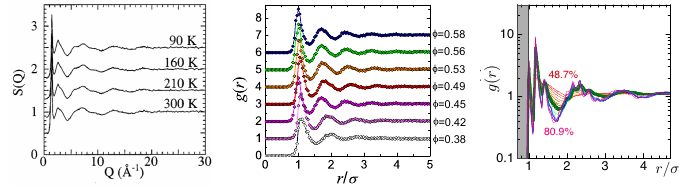}
\caption{\textbf{Two-point measures of structure.} (a) The static structure factor S(k) measured in propylene glycol, where $T_g=160$ K. Reproduced with permission from American Institute of Physics Copyright 1996 \cite{leheny1996}. 
(b) The radial distribution function $g(r)$ measured in a suspension of colloidal hard  spheres. Indicated is the effective volume fraction \cite{royall2018jcp}. 
(c) The radial distribution function $g(r)$ measured in a fluidized monolayer of bidisperse vibrated grains. Indicated is the effective volume fraction. The glass transition is marked by the thick green curve, taking place at a surface fraction of $0.744$. Reproduced with permission from American Physical Society~\cite{abate2006,keys2007}}
\label{figG}
\end{figure}

As stressed in the Motivation (section \ref{sectionMotivation}) and illustrated on Fig.~\ref{figG}, the microscopic structure of glassformers, when characterized by the pair correlation function, exhibits little change upon approaching the glass transition. Higher-order correlation functions may however better characterize the structural evolution in supercooled liquids approaching the glass transition. The role of local structure in the glass transition can be traced to the work of Frank~\cite{frank1952} who used energetic arguments to propose that icosahedral arrangements of atoms or molecules would be prevalent in supercooled liquids. Indeed, this is found to be the case for many glassformers where the interactions between the constituents are approximately spherically symmetric \cite{royall2015physrep}. Now, in atomistic glassformers (eg metallic glassformers), it is hard, though not impossible \cite{hirata2010,hirata2013,liu2013} to identify such structural motifs. Given that these geometric motifs are related to the interactions, and in (non--spherical) molecules, clusters have rather different topologies \cite{wales2004}, even if molecular coordinates were to be found, the geometric motifs formed would likely have a structure distinct to the ``canonical'' icosahedron. For a more extensive discussion, see~\cite{royall2015physrep}.

\subsection{Kirkwood $g_k$ in molecular liquids}
\label{sectionKirkwood}
In molecular liquids, owing to the experimental difficulties encountered when looking for higher-order structural motifs (such as icosahedra or even more complex motifs), most experiments have looked at \textit{two point} correlation functions, such as the structure factor $S(q)$. It turns out that $S(q)$ hardly evolves upon supercooling (Fig. \ref{figG}(a)) \cite{leheny1996}. Some more information about the correlations between the orientation of neighboring molecules can be gained by  measuring the Kirkwood factor $g_k$ which is the ratio between the measured static dielectric strength $\Delta \chi_1 = \epsilon(0)-\epsilon(\infty)$ and its expected value in an ideal gas of same density. Indeed, according to the standard interpretation $g_k = 1+z\langle \cos(\theta) \rangle$ where $z$ is the number of statically correlated molecules to a given molecule, and $\theta$ the angle between the central molecule and its neighbors.  However, recent theoretical progress~\cite{dejardin2022} have shown that this is a crude oversimplification, e.g. long range effects, as well as polarizability, are involved in $g_k$. Experimentally, $g_k$ hardly evolves upon supercooling \cite{nakanishi2011}, because the relative increase upon cooling is much smaller than that of the trivial factor $1/T$ (in an ideal gas $\Delta \chi_1 \propto 1/T$). Thus $g_k$ does probably not contain any structural information which could be decisive for the glass transition.

\subsection{Measuring higher--order structure in 3d colloidal glassfomers}

\begin{figure}[b!]
\centering
\includegraphics[width = 1.0\columnwidth]{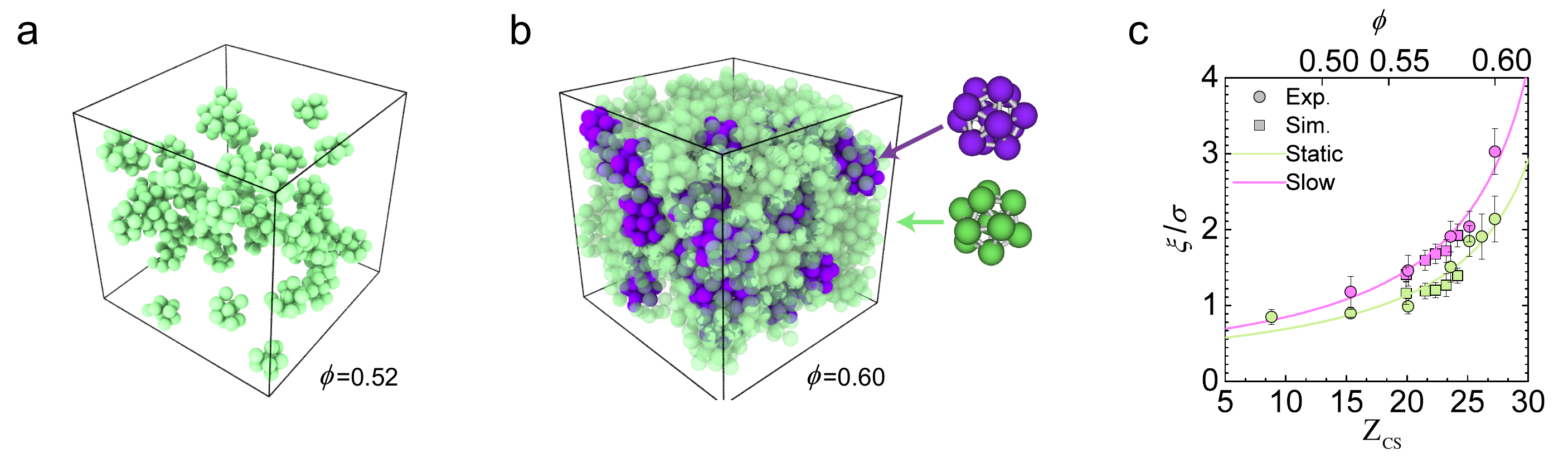}
\caption{\textbf{The change in structure in a colloidal glassformer and growth of structural and dynamical lengthscales}. Analysis of super--resolution images of small colloidal particles, allowing deeper supercooling than otherwise possible. 
(a,b) Renderings of locally favoured structures, 10--membered defective icosahedra (green) and 13--membered full icosahedra (purple) for volume fractions of $\phi=0.52$ (weakly supercooled) and $\phi=0.60$ (
deeply supercooled) respectively.
(c) Static and dynamic lengthscales $\xi_{\mathrm{Static}}$ and $\xi_{\mathrm{Slow}}$ obtained for defective icosahedra and slow moving particles respectively. Solid lines are fits inspired by RFOT. 
Reproduced from \cite{hallett2018}.
} 
\label{figJames}
\end{figure}

The first experimental study that clearly identified a change in structure in a colloidal system undergoing dynamical arrest in fact pertained to gelation~\cite{royall2008}. This was followed by work on a hard sphere like system which found a growth in \emph{both} fivefold symmetric order and locally crystalline order. The authors found that the local crystalline order was more strongly correlated with the slow dynamics than the fivefold symmetric order~\cite{leocmach2012}. Curiously, using \emph{the very same colloidal particles}, one of us found a growth in fivefold symmetry with little indication of crystalline order~\cite{royall2018jcp}. Whether this was due to small differences in the system (for example salt concentration and consequent electrostatic screening) or the different analysis methods used~\cite{steinhardt1983,malins2013tcc} remains unclear. Obtaining decisive conclusions obviously remains quite challenging. We remark that in~\cite{royall2008}, the \emph{topological cluster classification} \cite{malins2013tcc} algorithm was found to be more effective than bond--orientational order parameters in the case of gelation ~\cite{royall2008,steinhardt1983}.

The use of \emph{nano}--particle resolved studies (Section \ref{sectionReal}) 
enabled a regime of deeper supercooling to be explored. In particular, Fig.~\ref{figJames} shows a parallel growth of dynamical and icosahedral order lengthscales at deep supercooling, three decades in relaxation time past comparable studies. Further methods by which higher--order structure can be investigated in colloidal systems are reciprocal space techniques using sophisticated light scattering methods \cite{golde2016} such as X-ray cross--correlation \cite{wochner2009} and microbeam X-ray scattering \cite{liu2022}. Both of these techniques are highly promising, as they are not as limited in the size of the colloids that can be used, in the manner of real space microscopy, and a systematic analysis approaching the glass transition would likely result in considerable insight.

\subsection{Hexatic order in 2d colloidal systems and vibrated granular assemblies}
\label{sectionHexatic}

In 2d, the situation with local structure is profoundly changed. Unlike the case in 3d where fivefold symmetric structures found in the suoercooled liquid are incompatible with the crystalline order, in 2d the local packing is hexagonal, as is the crystal. Most experiments on mechanically shaken granular assemblies are conducted in 2d, so that the natural structural order parameter describes the so-called hexatic order, which captures the orientational order of the virtual bonds connecting the neighbouring grains or colloids. The order parameter reads:
\begin{equation}
    \Psi_6^j = \frac{1}{n_j}\sum_{k=1}^{n_j}e^{i6\theta_{jk}}
\end{equation}
where $n_j$ is the number of neighbors of $j$, the sum is made over the $n_j$ neighbors and $\theta_{jk}$ is the orientation of the bond $j-k$.
$\Psi_6$ is a complex number, whose modulus is $1$ if the neighbors are located at the vertices of a perfect hexagon and decreases to zero as the local disorder increases. Its phase indicates the local orientation of the hexagonal order.
\begin{figure}[b!]
\centering
\includegraphics[width =\columnwidth]{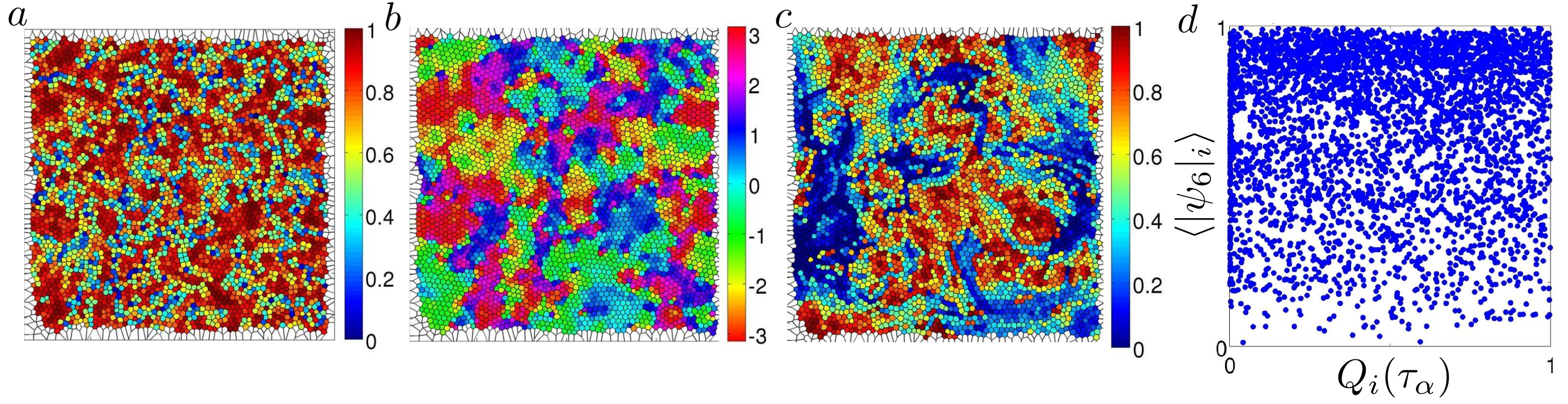}
\caption{{\bf  Comparison between maps of the bond-orientational parameter and the relaxation dynamics in a mechanically shaken granular assembly.} Adapted from Ref \cite{candelier2009phd}: (a) Modulus of $\psi_6$ averaged over $\tau_\alpha$. (b) Local orientation of the hexagonal order, phase of $\psi_6$, averaged over $\tau_\alpha$, color coded by the angle with the horizontal axis. (c) Local relaxation $Q_{i,t}(\tau_\alpha)$ induced by the displacements between $t$ and $t+\tau_\alpha$. (d) Scatter plot of the of $\left|\psi_6\right|_i$ averaged over $\tau_\alpha$ vs. $Q_i(\tau_\alpha)$; each point corresponds to one particle. 
}
\label{figGrainstruct}
\end{figure}

The influence of this local order on the dynamics has been debated in Refs  ~\cite{watanabe2008,sausset2008,kawasaki2008,tamborini2015}. In particular the presence of locally crystalline micro-domains, which may always be present in a 2d packing of discs  
may contribute to joint growth of the hexatic order and the relaxation time. The two signals are correlated by a common cause, but are not necessarily the cause of one another. Indeed there are examples where no correlations could be found~\cite{candelier2009}, as reported here, in Fig.~\ref{figGrainstruct} in the case of the mechanically shaken granular assemblies studied in~\cite{dauchot2005}.
Panels (a) and (b) respectively display the modulus and the phase of $\Psi^i_6$, while panel (c) displays the local relaxation $Q_{i,t}$. One hardly sees any similarities between the dynamical and the structural patterns, as confirmed by the scatter plot of the modulus of $\Psi^i_6$ vs. $Q_{i,t}$ in panel (d). The apparent contradiction in these results with others ~\cite{watanabe2008} surely merits further investigation.
Another anomalous example is flexible circular confinement, where the space at the walls due to the incommensurately between the hexagonal packing and the confinement in fact led to an acceleration in the dynamics for cases with high hexagonal symmetry~\cite{williams2015jcp}.

\subsection{The riddle of the different structural and dynamic lengthscales in molecules and colloids and grains}
\label{sectionRiddle}

Over the years, a considerable effort has been devoted to understanding structural and dynamical lengthscales in glassforming systems \cite{berthier2011,karmakar2014,royall2015physrep}. One observation that has emerged is exemplified here: \emph{molecular systems appear to have much smaller numbers of units (i.e. molecules) involved, despite the fact that they are much more deeply supercooled than colloids or grains} (or indeed dynamical data in computer simulation). Consider $N_\mathrm{corr}$ in Fig.~\ref{figNcorr} in which the molecular data reaches perhaps 10 at $T_g$ and only 2 at supercooling comparable to the colloidal data in Fig.~\ref{figJames}(c). The latter, however, has a \emph{lengthscale} of 2 or 3 \emph{diameters}, so in a 3d system this corresponds to 10-30 units, not the 2 in the molecular case. While these numbers are far from large (like all accessible lengths and volumes in equilibrated glassformers), they do not appear to be consistent. And the literature is filled with many more examples: colloids and grains (and computer simulation) typically seem to have many more units involved in relaxation than is the case for molecules \cite{dauchot2005,lechenault2008a,karmakar2014,royall2015physrep}.

We propose that the solution of the riddle lies in how we interpret the ``units''. For colloids (and grains) this is obvious -- a unit is a colloid or grain, as they are rigid bodies. For molecules the situation is more subtle: they are typically not rigid bodies and then one may consider the number of effectively rigid ``beads'' that are connected together in each molecule. This point is further discussed in ref.~\cite{stevenson2005} where RFOT predictions about thermodynamic-kinetic correlations were compared with available experimental data for molecular supercooled liquids. In short it turns out that 
typical molecules the number of beads per molecule is typically in the $3-7$ range \cite{stevenson2005}, (e.g. one obtains a value of $4.5$ for glycerol). Scaling the above numbers by 5 largely alleviates the discrepancy between the molecules and the colloids (and by inference, the simulations).\\

\section{Experimental evidence in support of the dynamical facilitation approach}
\label{sectionMu}

\begin{figure}[t!]
\centering
\includegraphics[width=120 mm]{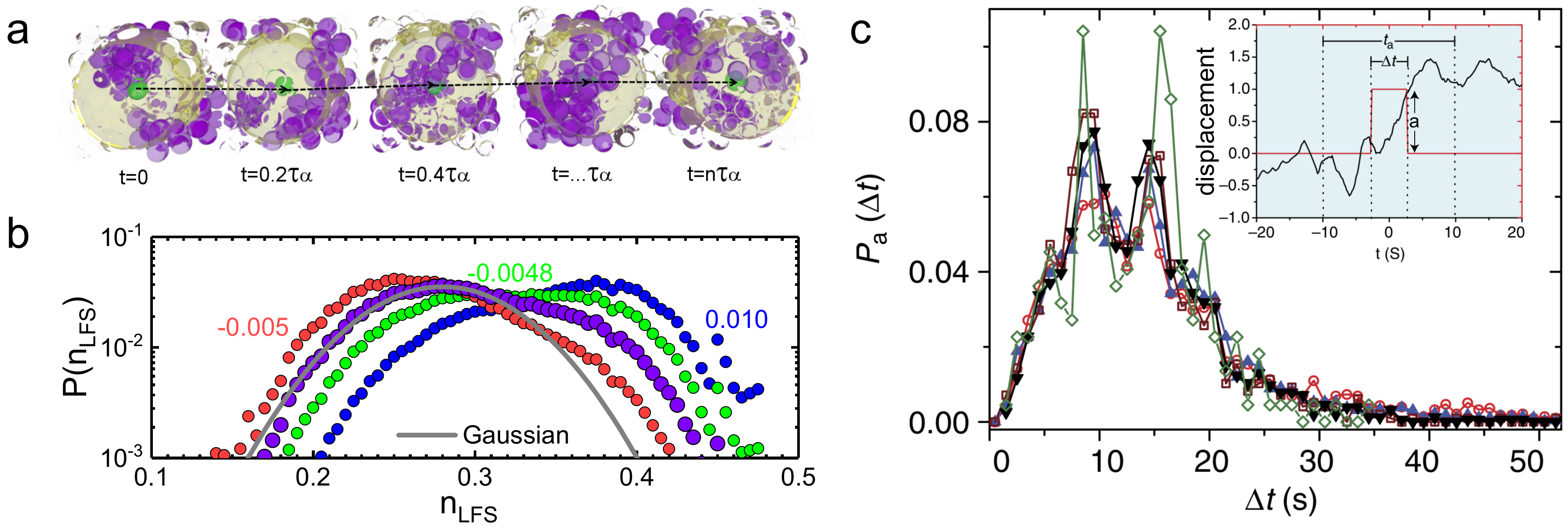}
\caption{\textbf{Testing the dynamic facilitation theory:} dynamical phase transition and excitations in colloidal experiments.
(a) The dynamical phase transition occurs in the space of \emph{trajectories}. Trajectories here are comprised of 100 particles closest to a central particle. Particles rendered in purple are in defective icosahedra, the locally favoured structure for this system. 
(b) Evidence for the dynamical phase transition is shown in the form of the probability distributions of populations of defective icosahedra for several biases ${\mu}$ at volume fractions ~$\phi=0.58$. The biasing field ${\mu}$ is analogous to chemical potential in conventional phase coexistence and here biasses towards a state with a high time-averaged population of defective icosahedra. Data modified from \cite{pinchaipat2017}.
(c) Identifying excitations which are the elementary units of relaxation in DF. Distribution of excitation times in a 2d colloidal system. $P(\Delta t)$ for $\phi=0.73$ (open red circles),$\phi = 0.74$ (filled blue triangles),  $\phi = 0.75$  (filled black inverted triangles), $\phi=0.77$ (open brown squares) and $\phi= 0.79$ (open green diamonds), showing that excitations are localized in time and do not change their character upon supercooling. (Inset) Representative trajectory of a particle in a excitation of duration indicated in red. The excitation time duration $\Delta t$ and the commitment time $t_a$ are marked by dotted lines. Data reproduced from~\cite{gokhale2014}.}
\label{figTrueMuHeart}
\end{figure}

The dynamical phase transition of dynamical facilitation (DF) theory was first identified in kinetically constrained models, which explicitly have no interesting thermodynamic properties in the sense that they are ideal gases. They nevertheless exhibit dynamical behaviour consistent with the systems we consider here, such as a massive increase in relaxation time and dynamic heterogeneity i.e. they are glassformers \cite{chandler2010}. The dynamical phase transition of DF occurs in \emph{trajectory space},in the form of two populations of trajectories, characterized by their \emph{activity} (akin to two populations of density in, e,g, a liquid--gas phase transition). 
Two such populations of trajectories, where trajectories are then small (sub)-systems of order 100 particles considered over a timescale comparable to or somewhat larger than the relaxation time $\tau_\alpha$, have been identified in colloidal systems. Here the activity is equated with the mean squared displacement of the particles in a trajectory \cite{pinchaipat2017,abou2018} with two populations of \emph{active} (normal liquid) and \emph{inactive} (glassy) trajectories upon biasing the trajectory population by suitable post-processing. Another way to identify the dynamic phase transition is through the population of locally favoured structures in trajectories, as indicated in Fig. \ref{figTrueMuHeart}(a,b)]. Here the active and inactive phases have small and large \emph{time-averaged} populations of LFS \cite{pinchaipat2017,royall2020jcp}.

Further studies of the dynamic facilitation theory include identifying the basic mechanism of relaxation, through \emph{excitations}, pockets of mobility where particles commit to a new position on a short timescale, as shown in Fig. \ref{figTrueMuHeart}(c) inset. As predicted by the theory, the nature of excitations remains essentially unchanged upon supercooling [Fig. \ref{figTrueMuHeart}(c)] \cite{gokhale2014}.\\

\section{Towards the configurational Entropy}
\label{sectionConfigurational}

\begin{figure}[t!]
\centering
\includegraphics[width=130 mm]{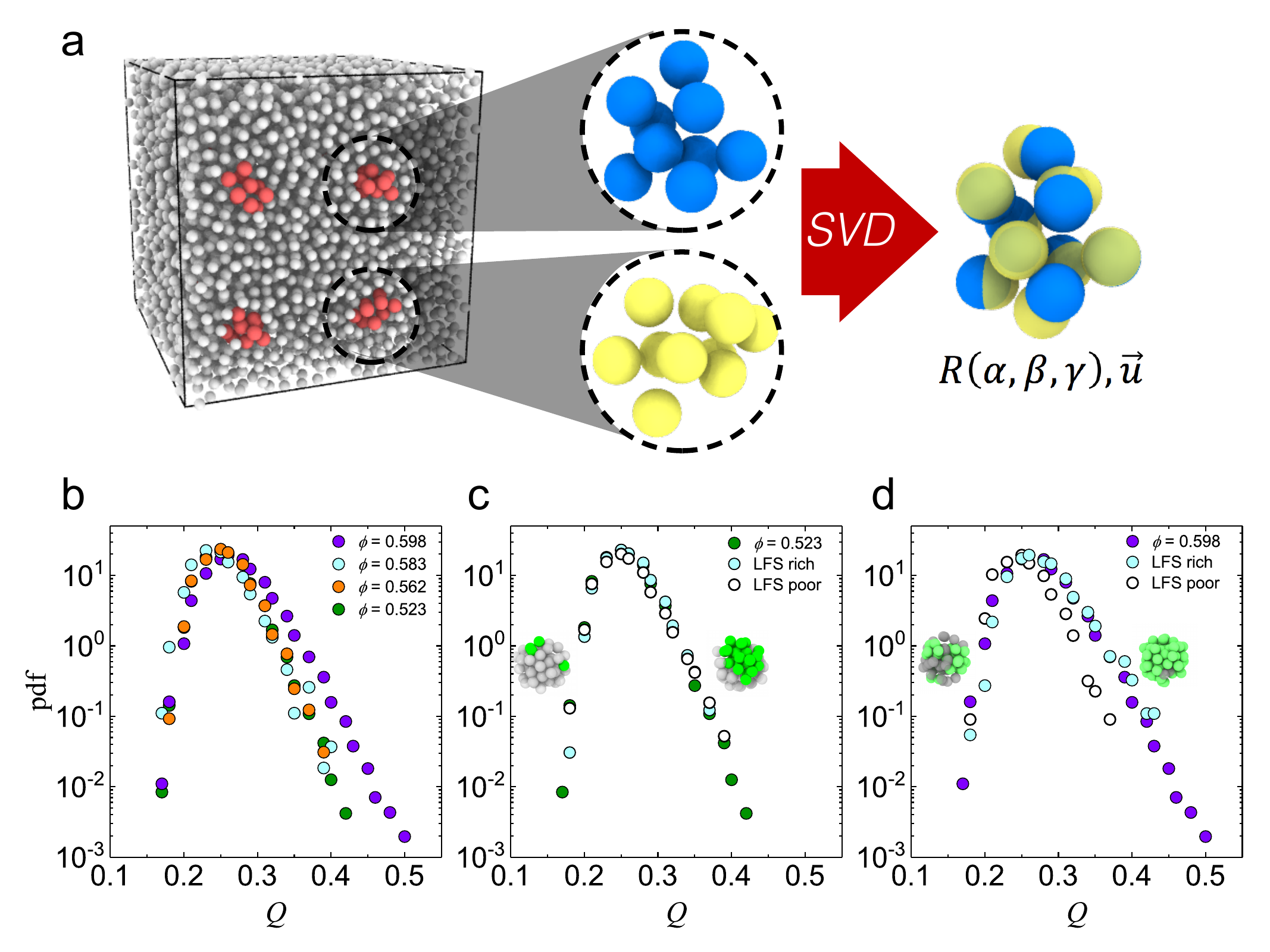}
\caption{\textbf{Evidence for a change in configurational entropy: Static overlap fluctuations in a colloidal glassformer.}
\textbf{a}, Schematic representation of the overlap calculation: Isolated clusters of $N_o$ particles are subsampled from a large configuration; for every pair of clusters, an optimal singular value decomposition (SVD) is found determining the rotation matrix $R(\alpha, \gamma, \beta)$ and the translation vector $\vec{u}$ that maximise the overlap. For ease of representation, here $N_o$=10 and Q=0.7. 
\textbf{b}, Probability distribution of overlap fluctuations for volume fractions 0.523, 0.562, 0.583 and 0.598. 
\textbf{c},\textbf{d}, Overlap fluctuations for all test spheres, and for the most LFS-poor (white) and rich (blue) 25\% test spheres, at volume fractions 0.523 (\textbf{c}, green) and 0.598 (\textbf{d}, purple). Shown in figures \textbf{c} and \textbf{d} are samples of LFS-poor (left) and rich (right) test spheres, where green particles are in defective icosahedra LFS and grey particles are not. Reproduced from \cite{hallett2018}.}
\label{figOverlap}
\end{figure}

The (configurational) entropy $S_\mathrm{c} (T)$ was originally determined in molecular systems, leading to a plot like that shown schematically in Fig. \ref{figCavagnaAngellHeartAll}(a), from where  \cite{tatsumi2012} the number $z(T)=S_c(\infty)/S_c(T)$ of molecules in a CRR is extracted, which yields $z(T_g)\simeq 4-6$, i.e. quite modest values. Note that here, $z(T)$ is regarded as a thermodyanmics, static quantity, distinct (though, depending on the theoretical standpoint, perhaps equal to) the number of units/beads making correlated motion $N_\mathrm{corr}$ discussed in sections \ref{sectionDynamical} and \ref{sectionRiddle}.

In colloidal systems, configurational entropy can be inferred via pinning a subset of particles \cite{williams2018} and was found to decrease upon approach to the glass transition. In unpinned systems insight can be gained from particle--based methods. In Fig. \ref{figOverlap}, we show the overlap of spherical regions of 64 particles in various locations in the system via a singular value decomposition. This yields a distribution of overlap [Fig. \ref{figOverlap}(b)], in which we see the emergence of a tail of high overlap at deep supercooling. We associate this tail of high overlap with a reduction in configuration entropy \cite{hallett2018}. Other methods to explore a reduction in configurational entropy in colloidal systems include \emph{pinning}, which may be induced by optical tweezers \cite{gokhale2014,gokhale2016jsm,gokhale2016} or adhesion of the colloids to a substrate \cite{williams2018}. 

Work to explore other predictions of the thermodynamic approach includes a measurement of the fractal dimension of so--called cooperatively rearranging regions (CRRs). This was found in colloidal systems to increase, consistent with more compact CRRs at deep supercooling~\cite{nagamanasa2015,mishra2015}. However equilibrating conventional colloidal systems past the mode--coupling crossover, where such compaction is expected \cite{berthier2011}, is hard. This can be addressed using small colloids, where indeed compaction of CRRs is found \cite{ortlieb2021}. The interface of CRRs is also predicted to have certain scaling properties \cite{biroli2017} and this too has been measured with colloidal systems in real space, where agreement was found with the predictions \cite{ganapathi2018}.

\section{Deeper in the glass}
\label{sectionGardner}

Until now, we have focused on equilibrated supercooled liquids approaching the glass transition. Here instead we turn our attention to the Gardner transition, which occurs in the (non--equilibrium) glass. When the cooling or compression rate is not slow enough, the system falls out of equilibrium, leaves the supercooled liquid branch (see coloured lines in Fig.~\ref{figCavagnaAngellHeartAll} and Fig.~\ref{figPatrickHeart}) and becomes a glass. Upon further compression, the hard sphere glass in infinite dimension undergoes a Gardner transition~\cite{gardner1985,kurchan2013,charbonneau2015}: glass states, envisioned as meta-basins in configuration space, break up in a hierarchy of marginally stable sub-basins at low enough temperature or high enough pressure. 

Eventually the hard sphere glass \emph{jams} at infinite pressure. The jamming transition is a geometrical transition taking place at zero temperature or infinite pressure, which controls the mechanical stability of athermal packings~\cite{van2009,liu2010,mari2009,berthier2009}. It occurs when particles, with well defined sizes and being compressed following a given protocol, cannot accommodate the imposed packing fraction without overlapping. It was shown that the exponents characterizing the criticality of the transition could be computed, following the compression of a given glass state to infinite pressure, provided that the presence of the Gardner phase is taken into account.
Conversely, there is experimental evidence in colloidal~\cite{zhang2009} and granular~\cite{coulais2012,coulais2014} systems that the diverging and vanishing length scales, which characterize the approach to the transition, survive in a finite, although modest, range of temperature for systems of soft spheres. 
The same holds true for the vibrational properties of the system~\cite{brito2010,chen2010,ghosh2010,kaya2010}, although extracting the latter from experimental data was shown to be a very delicate task~\cite{henkes2012}. In all cases, confirming the presence of a Gardner phase in deeply compressed glasses is a matter of great importance.
~\cite{charbonneau2017}.

The existence of a Gardner transition in finite dimension is now well established for hard sphere potentials, even if it is likely to be replaced by a crossover\cite{liao2019}. The same physics holds in a finite temperature range for softer potentials with a finite range~\cite{scalliet2019}. In the case of the soft Lennard Jones potential pertinent to molecules, there is not yet direct evidence of the Gardner marginality, but the stability of the glass and finite size effects may also hinder it.

\subsection{Experimental evidence for a Gardner crossover in granular glassformers}
\label{sectionDirect}
Mechanically shaken granular assemblies are particularly well--suited suited to identify an experimental signature of the Gardner transition. The grains interact via a very hard potential and the system is known to exhibit a jamming transition, where the mechanical pressure diverges, when the grains come into contact. Practically, their sizes allow for a direct visualisation not only of their positions, but also of their contacts. Crucially, the coordinates of the grains may be determined with much higher accuracy than for most colloidal systems, enabling very small changes in displacement to be measured reliably.

\begin{figure}[t!]
\centering
\includegraphics[width=0.8\columnwidth]{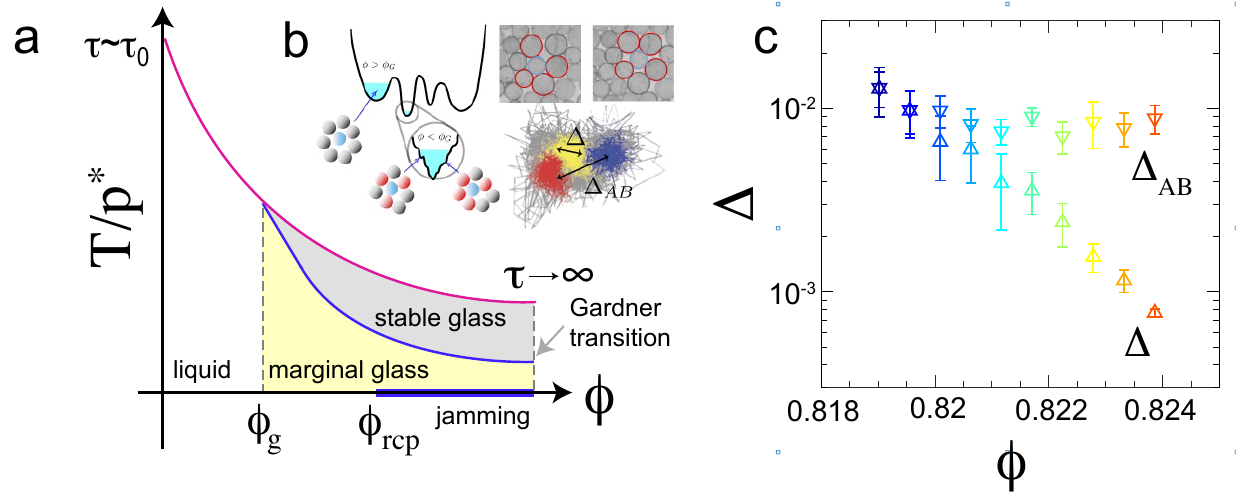}
\caption{\textbf{Evidence for a Gardner crossover in a mechanically shaken granular assembly} adapted from~\cite{seguin2016}. (a) The inverse pressure $1/p$ is shown as a function of the volume fraction $\phi$, with the pink line corresponding to the equilibrium equation of state.  At low volume fraction, the system is a fluid. For $\phi>\phi_g$ the system supports many meta-stable states with a range of pressures, which may be either stable glasses (dark shaded region) or marginal glasses (pale, yellow, shaded region) (based on \cite{charbonneau2014}).
(b) Upon crossing the Gardner transition, the energy landscape breaks into a fine structure showing ``energy minimina within minima within minima''.
Also shown is an experimental realization in a bi-disperse system of discs for two independent compression up to jamming within the same glass state.
(c) The cage size $\Delta(\phi)$ and the typical distance between the cages,$\Delta_{AB}$, obtained for different compression.}
\label{figPatrickHeart}
\end{figure}

Figure~\ref{figPatrickHeart} illustrates the real space realization of the full replica symmetry breaking taking place in a mechanically shaken granular assemblies. Once a highly compressed, jammed glass is obtained, the system is gently decompressed while ensuring that it remains in \emph{the same glass} and compressed again. This cycle is repeated many times for different target pressure. For sufficient compression, the final state differs from one compression to another:  in the glass phase $(\phi_g<\phi<\phi_G)$ the particle (in blue) is caged by its neighbors; those of which establish contact at jamming $(\phi_J)$ are not selected yet. In the Gardner phase $(\phi>\phi_G)$, each of the sub-basin eventually corresponds to a unique contact network (red neighbors). Below jamming, repeating 
compressions starting from the same glass state, where particles vibrate in a large (grey) cage, leads to a different caging location in a smaller cage (blue-red-yellow). The effect is quantified by comparing the average cage size within one state, $\Delta$, and the average distance separating the cages of the same particles across successive compression cycles, $\Delta_{AB}$. While for $\phi<\phi_G$,  $\Delta_{AB}$ decreases as $\Delta$, it plateaus to a constant value equal to $\Delta(\phi_G)$, when $\phi>\phi_G$, signing the entrance in the Gardner phase.

The transition to the Gardner phase should also be marked by large scale correlations in space and time. Quite remarkably those were reported experimentally~\cite{lechenault2008a,lechenault2010} \emph{before} the theoretical prediction of a Gardner phase was made. To do so the dynamics was probed at the scale of the contact, selecting a probe length $a$ smaller than a hundredth of the particle size, large dynamical heterogeneities were observed and a critical scaling was established for $G_4$.

\subsection{Experimental evidence for Gardner dynamics in a sedimenting  Brownian system}
\label{sectionMinute}
The direct observation reported in the previous section cannot be easily adapted to colloids due to the challenge of tracking the particles with sufficient accuracy \cite{ivlev2012}. An original method was elaborated in~\cite{hammond2020} in a system using particles $50\mathrm{\mu m}$ in diameter immersed in a solvent whose density is just below that of the particles. These particles undergo some Brownian motion, though the Brownian time to diffuse a diameter is of the order of a week! Nevertheless, they exhibit sufficient Brownian motion for the purpose intended and are large enough that their coordinates can be tracked accurately. As a result, starting from an initial disordered state, the particles slowly sediment, i.e. the number density of the system 
$\rho(t)$ continuously increases with the time $t$ elapsed since the initial  state. The system thus slowly tracks across the phase diagram. The key idea is to use the short time dependence of the Mean Square Displacement (MSD) of particles to probe the state that the system is experiencing. It turns out that, for $330$min $\le t \le 410$min, one observes $\mathrm{MSD}(\tau) \propto \ln(\tau/\tau^\star)$ where $\tau$ is the lag time over which the $\mathrm{MSD}$ is observed, and $\tau^\star \simeq 0.1$ms is the minimal lag time above which $\mathrm{MSD}$ can be measured. This logarithmic dependence, observed for $3$ decades in $\tau$, is interpreted~\cite{hammond2020} as a signature of the Gardner state that the systems experiences in the interval $330$min $\le t \le 410$min, before moving to another glass state at $t>410$min.

\subsection{A debated question in molecular liquids}
In molecular liquids, the first experiments looking for the Gardner transition were carried out in Ref. \cite{geirhos2018} by measuring  $\chi_{1}$ from $T_g$ to the lowest temperature ($5$K) in two liquids (namely sorbitol and xylitol) exhibiting a well-pronounced Johari-Goldstein $\beta$ relaxation. The excess of the measured $\chi_1$, with respect to what is expected in the $\beta$ relaxation regime, was related to a possible Gardner transition in these two liquids. Later, $\chi_3$ was measured \cite{albert2021}  
in glycerol below $T_g\simeq 188$K, down to $10$K. The motivation was that, in mean field ($d=\infty$), $\chi_3$ must diverge if a Gardner transition happens. As no sign of a divergence was detected for $\chi_3$, this was interpreted in Ref. \cite{albert2021} as revealing the absence of a Gardner transition in glycerol, at least for $T\ge 10$K. Because the liquids studied in Ref. \cite{geirhos2018} and in Ref. \cite{albert2021} are different (even though they are all poly-ols), and because in $d=3$ (as opposed to $d=\infty$), some Gardner-like physics might take place without implying a signature on $\chi_3$, we may regard the question of Gardner physics in molecular glasses as being still open.

\section{Discussion}
In this last section, we first recap what we have learnt from the work we have reviewed and then discuss what could be done to complement the present picture that we have for the glass transition. As is the case throughout this article, we focus on the complimentary work of the authors and co--workers.

\subsection{What have we learnt ?}
Despite the wild differences of their length-- and time-- scales, molecular, colloidal and granular glassforming systems appear to exhibit a universal slowing down at the macroscopic scale. If we are to assume that the same physics is responsible for this phenomenon, then we need to account for these differences in time-- and length--scales when considering these classes of glassformers and comparing between them. It therefore seems unlikely that a vibrated granular system can be equilibrated near $\phi_g$, by analogy to the relaxation time $\tau_\alpha(T_g)=100$s of molecular systems, due to the truly astronomical timescales involved (see Fig. \ref{figLengthscalesHeart}). In the case of colloids, however, there is a range of lengthscales and corresponding timescales. Some systems of clay particles (whose size can be $\sim$10nm) have been equilibrated for \emph{years} \cite{ruzicka2011}. Reference to Fig. \ref{figLengthscalesHeart} suggests that under these conditions equlibrating a colloidal system at $\phi_g$ ie $\tau_\alpha(\phi_g)/\tau_0=10^{14}$ might be realised.

Since this slowing down can be accounted for by theoretical approaches based on different principles, a tremendous effort has been carried out to refine our understanding of this slowing down, in particular to characterize the dynamics locally in space and in time. This led to the discovery of so-called dynamical heterogeneities, which are now well characterized in these three kinds of glassformers and have common universal features. However, this ``universality'' of the glass transition is, at present, far from being as strong as that of standard critical phenomena: firstly because the underlying cause of the dynamical arrest remains unresolved and secondly because demonstrating universal behaviour has some way to go. For example, the equivalent of locally favoured structures have yet to be identified in molecular systems, and the 2d vs 3d differences between vibrated granular systems and colloids/molecules suggests some differences in behaviour. Examples of these include the nature of the local structure (see section \ref{sectionStructure}) or even the nature of the glass transition itself \cite{berthier2019ncomms}.

\subsection{How could the "universal" picture be developed ?}
We have made the case for a universal picture of vitrification in the system classes that we have considered. We now consider how this might be developed.
In this spirit, we discuss below a subjectively chosen short list of questions, distinguishing between those pertaining mainly to dynamics and those related to structure.\\

\begin{itemize}

\item 
\emph{What are the absolute values for $N_\mathrm{corr}$ at the glass transition ($T_g,\phi_g$)}?
The measurement of lengthscales as we discuss above (section \ref{sectionRiddle}) seems to suggest that there is some evidence for a roughly fixed value for lengthscale of order 3--5 (blob) diameters at the glass transition $T_g$. We emphasize that, contrary to colloids and grains where real space imaging has been achieved, the absolute value of $N_\mathrm{corr}(T_g)$ is not precisely known in molecular glassformers (see section \ref{sectionDynamical}). This precludes a stringent test of existing theories, e.g. the RFOT prediction \cite{lubchenko2007} stating that $N_\mathrm{corr}(T_g) \simeq 195$ \textit{beads} for most liquids. Should $N_\mathrm{corr}(T_g)$ be much smaller than expected, this could even lead to a situation where some theories are not really distinguishable. We further emphasise that this considers the lengthscale associated with the Alpha relaxation time. In fact, recent computer simulation work \cite{ortlieb2021,scalliet2022} suggests that the lengthscale may depend on the timescale probed, with excitations of dynamic facilitation found at microscopic times and larger CRR--like entities at longer times. These observations are corroborated by experiments with colloids \cite{ortlieb2021,gokhale2014}. Obtaining a time--dependent lengthscale to investigate further these findings thus emerges as a major challenge in experiments with molecular glassforming systems.\\

\item 
\emph{Beyond the mere existence of Dynamical Heterogeneities, can we say more about the nature of the dynamical excitations, e.g. do they take the form of avalanches?} In the case of mechanically shaken grain assemblies, it was shown without ambiguity that dynamical heterogeneities result from a two time scale process~\cite{candelier2009}. On short time scales, clustered cage jumps concentrate most of the relaxation processes. On larger time scales but still rather shorter than that  corresponding to the mode--coupling crossover $\phi_\mathrm{mct}$, such clusters, akin to the cooperatively rearranging regions, aggregate both temporally and spatially in avalanches and ultimately build up the large scales dynamical heterogeneities. This was later confirmed for super-cooled liquids, performing molecular dynamics simulations on a two-dimensional model of glass-forming liquid and applying the same cluster analysis~\cite{candelier2010a}. This picture clearly suggests the presence of facilitation, although not necessarily conserved. However further examining experimental data from~\cite{keys2007}, it was shown that increasing the packing fraction of the granular assembly the number of CRR decreases, while their typical size increases and they appear more and more independently from each other : the form of facilitation responsible for the avalanche of CRR plays a less important role~\cite{candelier2010b}.

Conversely, recent computer simulations at deep supercooling have found that the elementary units of facilitation, so-called excitations, can become ``drowned out'' in the rapid relaxation in the mode-coupling regime ~\cite{ortlieb2021}. This is likely related to the dynamical spinodal nature of the mode-coupling crossover ~\cite{biroli2014} and the change in the nature of relaxation mechanism past the MC crossover ($T<T_\mathrm{mct},\phi>\phi_\mathrm{mct}$). At deep supercooling, simulations have also found multiple time relaxation processes ~\cite{ortlieb2021,scalliet2022}, with the emergence of compact clusters on short time scales. It also shares the same observation of an increasing size and a decreasing number of the initially relaxed clusters. The picture of avalanches obtained from the granular data at weaker supercooling ($\phi<\phi_\mathrm{mct}$) is now 
replaced by that of a slow coarsening of the CRR regions ~\cite{scalliet2022}. In this context facilitation is seen as playing a more important role at lower temperature ~\cite{ortlieb2021,scalliet2022}. Finding experimental verification of this multiple timescale--multiple lengthscale relaxation thus emerges as a key challenge.\\

\item 
\emph{Can we access higher order structural parameters in molecular systems and 3d granular experiments and indeed metallic glassformers \cite{liu2013,hirata2010,hirata2013}, similar to what has been achieved in recent years in colloids?} For molecular systems, we are not aware of decisive feature evidenced by macroscopic diffraction experiments. However one possible route forward is Reverse Monte Carlo, where two--point structure is matched to computer simulation and the resulting higher--order simulated structure analysed \cite{mcgreevy2001}, which is used to considerable effect in some other classes of glassforming system \cite{salmon2013}. As for local investigations (at the nano--scale), it seems hard
to avoid that any probe which is molecular in size will change the interactions felt by the glass forming molecules probed by the experiment: this is already known from the changes in $T_g$ observed in polymers near free surfaces, or by the extremely involved variations measured when inserting glass forming liquids in nano-porous media. For $3d$ granular systems, the situation seems less challenging but a dedicated setup remains to be conceived. One possibility is to make use of hydrogel spheres in index matched scanning experiments as initiated in~\cite{dijksman2017}.\\

\item 
\emph{Can an equivalent of $S_\mathrm{conf}$ be measured in granular systems?} This would allow one to test more deeply to which extent one can rely on an effective thermodynamics for granular systems. A starting point could be to try to adapt the method recently conceived for colloids (see section \ref{sectionConfigurational}).  \\

\item
\emph{Can we develop further measurements of the Point to Set length $L_\mathrm{pts}$ length?} The $L_\mathrm{pts}$ length-scale is a central concept for thermodynamic theories \cite{lubchenko2007,tarjus2005} of the glass transition since it gauges the influence that (amorphous) boundary conditions have on local thermodynamics. Establishing that $L_\mathrm{pts}$ exists and grows upon cooling would be a decisive argument in the controversy between the thermodynamic and the dynamic scenario for the glass transition. In the first numerical study \cite{biroli2008} $L_\mathrm{pts}$ was studied by using a well equilibrated configuration that is suddenly frozen outside a cavity of size $L$. It was shown later \cite{cammarota2012} that this is a particular case of the more general ``random pinning'' idea where, by freezing a fraction $c \propto 1/L_\mathrm{pts}^3$ of the particles randomly chosen in the system, one induces, for $c \ge c_\mathrm{critital}(T)$ a phase transition towards an ideal glass state. predicted by RFOT. To address this question experimentally, it seems extremely challenging to perform in a molecular glass an experiment similar to that achieved in colloids \cite{hima2015,gokhale2016,williams2018}: we are aware of two preprints \cite{das2021,kikumoto2021}  where a concentration $c$ of large molecules, dispersed in short molecules, is considered as a pinning field, and where the changes in the dynamics is interpreted as revealing the pinning effect associated to $L_\mathrm{pts}$ (see however an alternative interpretation proposed in \cite{kikumoto2021} where the small molecules are seen as plasticizers of the large ones). As any experiment in molecular glassformers will probably turn out to be a very non ideal realization of pinning, we foresee that \textit{several} kinds of experimental techniques will be needed to obtain a consensus on the existence of $L_\mathrm{pts}$, just as what happened two decades ago, when several experimental \emph{tour de force} were needed to conclude about the existence of dynamical heterogeneities in supercooled molecular liquids (see Ref. \cite{richert2011} for a review). This will, surely, demand a real theoretical effort to adapt the pinning idea to the realm of these future experimental breakthroughs.\\

\item \emph{Is it possible to obtain experimental evidence for a dynamical phase transition between active and inactive trajectories in grains and molecules, as has been demonstrated in the case of colloids (section \ref{sectionDynamical})?} In the case of vibrated grains, one imagines that an approach similar to that employed for the colloids might be appropriate. For molecules, less direct methods may need to be employed, we note for example the intriguing recent results of Jin \emph{et al} \cite{jin2021}.\\

\item \emph{Another theory of the glass transition is Geometric Frustration, which posits an avoided transition in a curved space \cite{tarjus2005}}. While a 3d curved space is virtually impossible in experiment, elegant work with a 2d hyperbolic curved space has been carried out in computer simulation \cite{sausset2008}. Experiments with colloidal systems have been carried out which exhibit such hyperbolic curved space  \cite{irvine2012} and thus have the potential for realising the intriguing results of the simulations.

\end{itemize}

%

\section*{Acknowledgements}

The authors wish first to thank the hundreds of authors of the papers cited in this review for their contribution to our understanding of the glass transition. Besides, we wish to thank  those with whom we have been fortunate to work with in the context our  work which is cited here :
Christiane Alba-Simionesco, Samuel Albert,
Paul Bartlett, Ludovic Berthier, 
Giulio Biroli, 
Jean-Philippe Bouchaud, 
Coralie Brun, 
Chiara Cammarota, Matteo Campo,
Luca Cipeletti, 
Caroline Crauste-Thibierge, 
Pierre-Michel Déjardin, 
Paramesh Gadige, 
James Hallett, 
Nathan Israeloff,
Walter Kob, 
Alois Loidl, Peter Lunkenheimer,
Alex Malins, 
Levke Ortlieb, 
Erdal O\u{g}uz,
Takehiro Ohtsuka,
Rattachai Pinchipat,
Ranko Richert,
Camille Scaliett, 
Thomas Speck, 
Hajime Tanaka, 
Gilles Tarjus, 
Francesco Turci, 
Pierfrancesco Urbani,  
Ian Williams, 
Stephen Williams
for many insightful discussions over the years. CPR wishes to acknowledge European Research Council (ERC Consolidator Grant NANOPRS, project number 617266), the ANR through projects DiViNew and the Royal Society. FL acknowledges the support of ANR through project COMET.

\bibliographystyle{crunsrt}
\nocite{*}
\bibliography{heart.bib}

\end{document}